\documentclass[mathleft]{an}
\usepackage{graphicx,lscape}
\usepackage{times}
\begin{document}
\Pagespan{1}{21}
\Yearpublication{2014}%
\Yearsubmission{2014}%
\Month{4}%
\Volume{1}%
\Issue{1}%

\title{A direct imaging search for close stellar and sub-stellar companions\\ to young nearby stars\thanks{Based on observations made with ESO telescopes at Paranal Observatory under programme IDs 083.C-0150(B), 084.C-0364(A), 084.C-0364(B), 084.C-0364(C), 086.C-0600(A) and 086.C-0600(B).}}

\author{Nikolaus~Vogt\inst{1}\thanks{Corresponding author: \email{nikolaus.vogt@uv.cl}} \and Markus Mugrauer\inst{2} \and Ralph Neuh\"{a}user\inst{2} \and Tobias O.B. Schmidt\inst{2} \and \\ Alexander Contreras-Quijada\inst{1} \and J\'{a}nos~G.~Schmidt\inst{2}}

\titlerunning{A direct imaging search for close stellar and sub-stellar companions to young nearby stars}
\authorrunning{Vogt et al.}

\institute{Instituto de F\'{i}sica y Astronom\'{i}a, Universidad de Valpara\'{i}so, Chile, Avenida Gran Breta\~na 1111, Valpara\'{i}so, Chile;
\and Astrophysikalisches Institut und Universit\"{a}ts-Sternwarte Jena, Schillerg\"{a}{\ss}chen 2, D-07745 Jena, Germany}

\received{-}
\accepted{-}
\publonline{later}


\abstract{A total of 28 young nearby stars (ages $\leq 60$\,Myr) have been observed in the K$_{\rm s}$-band with the adaptive optics imager Naos-Conica of the Very Large Telescope at the Paranal Observatory in Chile. Among the targets are ten visual binaries and one triple system at distances between 10 and 130 pc, all previously known. During a first observing epoch a total of 20 faint stellar or sub-stellar companion-candidates were detected around seven of the targets. These fields, as well as most of the stellar binaries, were re-observed with the same instrument during a second epoch, about one year later. We present the astrometric observations of all binaries. Their analysis revealed that all stellar binaries are co-moving. In two cases (HD\,119022\,AB and FG\,Aqr\,B/C) indications for significant orbital motions were found. However, all sub-stellar companion-candidates turned out to be non-moving background objects except PZ Tel which is part of this project but whose results were published elsewhere. Detection limits were determined for all targets, and limiting masses were derived adopting three different age values; they turn out to be less than 10 Jupiter masses in most cases, well below the brown dwarf mass range. The fraction of stellar multiplicity and of the sub-stellar companion occurrence in the star forming regions in Chamaeleon are compared to the statistics of our search, and possible reasons for the observed differences are discussed.}

\maketitle

\section{Introduction}

Since 1995, many planet-candidates were discovered with the radial velocity technique (RV), which gives the lower minimum-mass limit ($m \cdot sin(i)$), some of them confirmed by transit or astrometry. One important aspect is the brown dwarf desert (Grether \& Lineweaver 2006\nocite{grether2006}), i.e. missing companions with a minimum-mass between 20 to 50 Jupiter masses ($M_{Jup}$). Only $\leq 1\,\%$ of the observed solar mass stars have such close brown dwarf companions, even though they would be easily detectable with RV-methods. Most known RV-planet-candidates and transiting planets have very small separations, below $\sim 6$\,AU,  mainly due to observational biases (sensitivity) and the relatively short epoch difference since the starting date of applying these techniques. The interesting open question is now: Are there brown dwarf companions at intermediate separations, i.e. between $\sim 6$\,AU and tens of AU. Instead of waiting a few decades until the RV- and transit-technique cover such long orbits, we used the adaptive optics imager Naos-Conica (NACO, see Lenzen et al. 2003 and Rousset et al. 2003\nocite{lenzen2003}\nocite{rousset2003}) of the Very Large Telescope (VLT) at the Paranal Observatory in Chile, in order to search for such companions with direct imaging at the highest possible angular resolution.

Our working group (a collaboration between the Astrophysical Institute and University Observatory in Jena, Germany and the Instituto de F\'{i}sica y Astronom\'{i}a in Valparaiso, Chile) is applying this method since several years, and has already detected various sub-stellar companions of young stars, most of them in star-forming regions, in particular GQ\,Lup (Neuh\"{a}user et al. 2005, 2008\nocite{neuhaeuser2005}), CT\,Cha (Schmidt et al. 2008\nocite{schmidt2008a}), Cha\,H$\alpha$\,2 (Schmidt et al. 2008b\nocite{schmidt2008b}), TWA-5 (Neuh\"{a}user et al. 2000a, 2010\nocite{neuhaeuser2010}) and PZ\,Tel (Mugrauer et al. 2010, 2012; Schmidt et al. 2014). Stellar companions seem to be more frequent than sub-stellar ones, as shown in the star-forming region Chamaeleon (Vogt et al. 2012; Schmidt et al. 2013\nocite{vogt2012}\nocite{schmidt2013}). A preliminary report on our results, concerning stellar and sub-stellar companions in the star-forming region Lupus is given by Vogt et al. (2013). In the present paper we apply the same method, searching the surroundings of a total of 28 selected young, nearby field stars, in order to elucidate the occurrence and improve the statistics of recently formed binaries or multiple systems with low-mass stellar or sub-stellar companions, and to compare the  results with those of star-forming regions. In Section 2 we describe the target selection criteria and give some details about our observing and reduction procedure. Section 3 describes the detection limits of our observations. In Section 4 we present our astrometry results of binaries and of those cases with detected companion-candidates, while Section 5 contains a discussion and conclusions.

\section{Target selection, observations}

Our targets should be young, with possible companions still self-luminous in infrared, and nearby in order to reach the absolutely faintest sub-stellar companion-candidates and to reach low masses. In addition, for nearby stars we know the most precise parallax-values, as measured by the HIPPARCOS-satellite (van Leeuwen 2007\nocite{leeuwen2007}). To reach projected separations of a few to tens of AU, the targets need to be located at distances of the order of 100\,pc or closer. Our age-limit is about 60\,Myr, ensured with the presence of moderately strong Lithium absorption at 6708\,\AA. The targets also need to have sufficiently large proper motions ($\geq 40$\,mas/yr, i.e. few NACO pixels per year) for a significant common proper motion check after about one year. Finally the selected stars should be observable from the southern hemisphere, and no deep astrometric quality images should have been published (or be present in the public ESO archive) at the beginning of our campaign. A selection according to these criteria led to a list of about 100 targets, 29 of them well observable within the period September to March. Table\,\ref{table_targets} lists some of the published information on the 28 targets selected for this study (the 29$^{th}$ case is PZ\,Tel, see Section 5). For eight targets the ages are missing because their errors from H-R D and isochrones are very large, thus the age values are unconstrained; but also these targets are probably young due to Lithium absorption.

The birth line in our age values is defined such that the young stars appear as optically visible objects for the first time; the pre-main sequence isochrone is zero at that time. As stated above, our main age criterion was the presence of Lithium absorption at 6708\,\AA. Two of our targets, HD 6665 and HD 217352 (with moderately strong Lithium abundances according to Strassmeier et al. 2000)  later turned out to be post-main sequence giants with Lithium enrichment attributed to the conversion of $^{3}$He via $ ^{7} $Be to $ ^{7} $Li by the Cameron-Fowler mechanism (Kumar et al. 2011). We include them here for completeness, as they were also observed during our campaign. We also searched the literature for additional age values determined by other methods and list them in the last column of Table 1.

We used NACO with its S13 objective, which exhibits a pixel-scale of about 13\,mas/pixel and a field of view of about 13.5\,arcsec$\times$13.5\,arcsec. All targets were observed in the K$_{\rm s}$-band with typical total exposure times of 20 to 30 minutes per target, composed of many short integrations of only 0.347\,s, the shortest possible integration time (DIT) with NACO in full frame readout-mode, in order to minimize saturation effects, induced by the bright target stars. In order to remove the bright sky-background in the infrared the jitter-technique from the eclipse pipeline (Devillard 1997) was applied, using a random jitter-width of 7\,arcsec. The first epoch observations were obtained on 2009, September 28-30, and December 7-8 and 2010, February 22-23. A second epoch was only taken for those targets with companion-candidates detected during the first epoch observations, and for some of the other stellar binaries to follow their orbital motion. This second epoch refers to observations on 2010, October 24-29 and on 2011, March 24-25.

Our observing and reduction strategy, the calibration-procedures and the astrometric data analysis, applied in general by our working group, are described in more detail by Vogt et al. (2012). Some additional information referring to the particular observing runs discussed here can be found in Mugrauer et al. (2010). The NACO detector was astrometrically calibrated in each observing epoch with images of the globular cluster 47\,Tuc. The achieved astrometric calibration for all observing epochs, whose data are presented in this paper, are summarized in Table\,\ref{table_astrocal}.

\section{Detection limits and masses of detectable sub-stellar companions}

The achievable detection limit depends on the observing conditions (atmospheric conditions, i.e seeing and coherence time), as well as the total observing time, but also on the angular separation and magnitude difference from the bright target star. We have determined these limits for all 28 targets, which are discussed here. Four typical examples of the dynamical ranges vs. angular separation are shown in Figure 1. Table 3 gives the corresponding numerical values $\Delta K_{s}$ for the separations 0.2, 0.5 1.0 and 1.8 arc seconds of all targets. The ($S/N=3$) detection limits reached in the background limited region, i.e. at larger angular separations typically of more than 1.5\,arcsec around the bright central stars, are given for all targets in Table 4. This table also lists the number of obtained observing epochs (N$_{ep}$) of all targets, as well as the number of detected companion-candidates (N$_{cc}$). In addition, the minimum-masses of detectable sub-stellar companions for three different ages, namely 10, 50, and 100\,Myrs, are listed. The minimum-masses were derived with the measured detection limits in the K$_{\rm s}$-band, the known parallaxes of the stars (from HIP, see van Leeuwen 2007\nocite{leeuwen2007}), the assumed ages, as well as the Baraffe et al. (2003) evolutionary models. The same models were used to determine the projected minimum-separations (in AU), beyond which all stellar companions (mass $>75$\,M$_{\rm Jup}$) of the observed targets can be detected in our deep NACO imaging data, using the measured dynamic ranges, the known K$_{\rm s}$-band magnitudes of all targets (from the 2MASS Point Source catalog, see Skrutskie et al. 2006\nocite{skrutskie2006}), as well as their parallaxes. The minimum-separations were again derived for the three assumed system ages mentioned above.

\section{Astrometry results}

All stellar binary or triple systems listed in Table 1 were known previously, they are listed in the Washington Visual Double Star Catalog (Mason et al. 2001\nocite{mason2001}), some with rather remote first epochs back to the 19$^{th}$ century. Since the aim of this paper is not the analysis of their orbits, we give in Table 5 the projected separations and position angles measured by us in our NACO images of these systems, together with the epochs of observation. In Fig.\,\ref{figure_binastro} we show the proper motion analysis for those stellar systems, observed by us in two observing epochs. These diagrams confirm that all five systems are clearly co-moving binaries. Due to the small time difference between our two NACO observing epochs of the order of one year, we could not find clear indications of orbital motion in most cases. Only the close binaries HD\,119022\,AB and FG\,Aqr\,B/C reveal significant orbital motions in both, angular separation and position angle. Marginal orbital motion may be present in the A/B pair of the triple system FG\,Aqr (in position angle) and in HD\,47787 (in angular separation). We can constrain their orbits as follows: since HD 47787 does not show significant change in PA, its orbit cannot be approximately circular pole-on; we do see a marginally significant change (1$\sigma$) in separation, so that the orbit is approximately edge-on and/or eccentric. A similar conclusion can be drawn for both HD 199022 A+B and FG Aqr B+C: In these pairs, both PA and separation change significantly, so that these orbits are either eccentric and/or the inclination is between pole-on and edge-on. 

In a total of seven target fields we detected faint companion-candidates, in five of them only one per field, but around HD\,283798 ($b=-13^{\circ}$) three and around HD\,139084 ($b=-2^{\circ}$) even twelve possible companion-candidates. All these fields were observed in two epochs, the astrometric results are listed in Table 6. In Figure 3 we present the corresponding proper motion analysis. It turns out that none of the 20 candidates seem to be gravitationally bound to the target stars; all are compatible with background sources, i.e. they exhibit only small or negligible proper motions. However, our measurements are not accurate enough to reject definitely the companion hypothesis in the case HD283798 with its three companion candidates, but their astrometry is more consistent with none-moving background objects than with bound co-moving companions. However, our NACO astrometry cannot confirm this background hypothesis significantly. A further follow-up observation of these companion-candidates is needed.

The large number of companion candidates around HD 139084 and HD 283798 are well compatible with the rather low galactic latitude of them, compared to most other targets, which are located far away from the galactic plane.

We believe that it is important to publish also negative results in order to avoid that future observers will waste valuable observing time at large telescopes  using our results as a starting point for their studies. Therefore, we give here also the limiting magnitudes and details of the minimal angular separations for each target, depending on the $ \Delta $K$_{s}$ magnitude difference.  In addition, our results will be important for future statistics on the frequency of low-mass companions to young stars.

\section{Discussion and conclusions}

In addition to the 28 targets presented here, our search program also contained PZ\,Tel, a member of the approximately 12\,Myr old $\beta$\,Pic moving group. In this case, a faint co-moving companion separated by only about 0.3\,arcsec from its host star PZ\,Tel\,A could be imaged at first with NACO at the VLT in 2007, as described in detail by Mugrauer et al. (2010). The JHK photometry of the companion, PZ\,Tel\,B, is consistent with a brown dwarf of a mass between 24 and 40\,M$_{\rm Jup}$. Additional follow-up observations of PZ\,Tel\,B (Mugrauer et al. 2012\nocite{mugrauer2012}) confirmed the significant orbital motion of the companion around its host star, constraining its orbital parameters. Apparently the system is seen almost edge-on and characterized by a high eccentricity.  Spectroscopic observations of Schmidt et al. (2014) confirm the brown dwarf nature of PZ\,Tel\,B, revealing an effective temperature of about 2500\,K and a luminosity L$_{bol}=0.0025\,L_{\odot}$. They compared these results with evolutionary models and derived a best-fitting mass of about 21\,$M_{Jup}$, at a lithium depletion age of 7\,Myr.

Is it surprising that no other new co-moving stellar or sub-stellar companion was found among the remaining 28 targets presented here?

To answer this question we can compare the result of our study with results of similar searches in several nearby star-forming regions. The multiplicity fraction with stellar companions (detected also by direct imaging, i.e. in a similar range of separations and magnitude difference as in our study) ranges typically between 28 and 36\,\%, as shown in the star-forming regions Lupus (Ghez et al. 1997\nocite{ghez1997}), Scorpius-Centaurus (K\"{o}hler et al. 2000\nocite{koehler2000}), Ophiuchus (Ratzka et al. 2005\nocite{ratzka2005}), Corona Australis (Neuh\"{a}user et al 2000b; K\"{o}hler et al. 2008\nocite{koehler2008}) and Chamaeleon (Lafreni\`ere et al. 2008; Vogt et al. 2012; Schmidt et al. 2013\nocite{lafreniere2008}\nocite{vogt2012}\nocite{schmidt2013}). In our sample of a total of 28 stellar systems there are ten stellar binaries (as summarized in Table 5 plus the two wide binaries HD\,13246\,AB and HD\,33802\,AB, whose components were observed individually in this study) and one stellar triple system, hence our multiplicity fraction is $11/28 = 39\,\%$ ($\pm 19\,\%$), similar to the typical values for star-forming regions.

Concerning the frequency of sub-stellar companions, i.e. brown dwarfs or planets, we can consider the situation in the Chamaeleon star forming region: There are three certain sub-stellar companions: CHXR\,73 (Luhman et al. 2006\nocite{luhman2006}), Cha\,H$\alpha$\,8 (Joergens \& M\"{u}ller 2007\nocite{joergens2007}) and CT\,Cha (Schmidt et al. 2008a\nocite{schmidt2008a}). In addition, there are four candidates at the brown dwarf-star limit: Cha\,H$\alpha$\,2 (Schmidt et al. 2008b\nocite{schmidt2008b}), T\,Cha (Hu\'elamo et al. 2011\nocite{huelamo2011}), DI\,Cha, as well as Cha\,H$\alpha$\,5 (Schmidt et al. 2013\nocite{schmidt2013}). In Chamaeleon a total of 46 targets have been investigated with the same methods, applied here. This leads to about 7\,\% ($\pm 15\,\%$) of Chamaeleon members with certain sub-stellar companions and another 9\,\% of brown dwarfs candidates.

According to these statistics we would expect one certain sub-stellar companion but two additional candidates at the stellar/brown dwarf mass-limit among all our observed targets. Only one of three expected possibilities happened (PZ\,Tel\,B). The reason for this could be, of course, just accidental, due to the statistics of small numbers. Another reason could be the fact that most of our targets are older than those in the Chamaeleon and Lupus star-forming regions (with 2 to 10\,Myrs, in average). On the other hand, all targets of this study are much closer than those in the star-forming regions in Chamaeleon and Lupus, whose distances are of the order of 140 to 165\,pc. Therefore, most of our companion mass-limits, given in Table 4, reach well down to the typical planetary mass range, and therefore also cover the entire mass range of brown dwarfs. If there is any brown dwarf companion around one of our targets (within the NACO field of view, and not too close to its host star), we should have detected it. 

If companions are distributed equally in three-dimensional space around the primary, and since we always observe with direct imaging the two-dimensional projection (on sky), we can detect about two thirds of the wide companions, those separated well from the primary in right ascension and/or declination (or any two other directions on sky), but not in the third dimension (before or behind the primary). In the third dimension, we can detect neither wide nor close companions. The detection rate of direct imaging is therefore 2/3 of all companions outside a (close-in) separation limit. A similar completeness ratio of 62$  \%$ was determined by Metchev \& Hillenbrand (2009), based on brown dwarf observations in the Palomar/Keck Adaptive Optics Survey whose observing conditions were similar to those of our VLT/NACO instrument.

Hence, a further possibility for the missing brown dwarf companions might be that they are currently located close to their host stars, passing through their pericenters and therefore remain undetectable for direct imaging searches. As in the case of PZ\,Tel\,B, these companions could then be identified in a couple of years via direct imaging observations. Final statistical conclusions have to await the results of our data in Chamaeleon, Lupus, Corona Australis and other star-forming regions, which are under analysis at present.

\acknowledgements
We would like to thank the ESO Paranal Team, the ESO User Support department and all other helpful ESO services. We made use of the Simbad and VizieR databases operated at CDS in Strasbourg, France. N.V. acknowledges support by the Gemini-CONICYT fund 32090027 and DIUV 38/2011 fund. M.M. acknowledges the Deutsche Forschungsgemeinschaft (DFG) for support in program MU2695/13-1.

\newpage
\clearpage

\begin{landscape}
\begin{table}
\centering
\caption{The list of all observed targets with their names, equatorial and galactic coordinates, spectral types (SpT), and K$_{\rm s}$-band magnitudes, as listen in the Simbad database, sorted by increasing right ascension. Parallaxes ($\pi$) are from HIPPARCOS measurements (van Leeuwen 2007\nocite{leeuwen2007}). The 6th column refers to the known multiplicity status of the targets with 1 for single, 2 for binary, and 3 for a triple system. Columns 9 and 10-11 give the masses and ages of all targets, according to Tetzlaff et al. (2011) and for additional stars obtained with the same technique. The uncertainties of the masses are 0.1\,M$_{\odot}$ unless stated differently. Ages  from other sources are listed in the last column.\newline}
\begin{tabular}{ccccccccccc}
\hline
Target & $\alpha $ (2000) & $\delta $ (2000) & l,b & $\pi $  & Multi- & SpT & K$_{\rm s}$ & Mass & \multicolumn{2}{c}{Age [Myr]} \\
 &  [hh mm ss] & [dd mm ss] & [$^{\circ}$, $^{\circ}$] & [mas] & plicity &   &  [mag] & [$M_{\odot}] $ & Tetzlaff et al.(2011) & Other sources\\
\hline
HD\,225213        & 00 05 24.4 & $-$37 21 26 & 344,$-$76 & $230.4 \pm 0.9$ & 1  & M1.5V & 4.5 & $5 \pm 4$     & $ \leq $ 50 & \\
HD\,105           & 00 05 52.5 & $-$41 45 11 & 333,$-$72 & $25.4 \pm 0.6$  & 1  & G0V   & 6.1 & 1.1           & $26.7 \pm 4.2$ & 30(2)\\
HD\,6665          & 01 07 36.4 & $+$20 33 43 & 128,$-$42 & $4.5 \pm 0.8$   & 1  & G5V   & 5.5 & 1.2 & & Post-MS(1) \\
BD+17\,232        & 01 37 39.4 & $+$18 35 33 & 138,$-$43 & $15.6 \pm 2.0$  & 2  & K3Ve  & 6.7 & 0.9               & & \\
HD\,8558          & 01 23 21.3 & $-$57 28 51 & 295,$-$59 & $20.2 \pm 0.8$  & 1  & G7V   & 6.8 & 1.0           & $28.3 \pm 8.9$ & \\
HD\,9054          & 01 28 08.7 & $-$52 38 19 & 290,$-$64 & $27.8 \pm 1.0$  & 1  & K1V   & 6.8 & 0.8           & $54.6 \pm 8.2$ & \\
HD\,13183         & 02 07 18.1 & $-$53 11 57 & 280,$-$60 & $19.6 \pm 0.6$  & 1  & G7V   & 6.9 & 1.0           & $32.3 \pm 7.9$ & \\
HD\,13246\,A      & 02 07 26.1 & $-$59 40 48 & 286,$-$55 & $22.6 \pm 0.5$  & 1  & F7V   & 6.2 & 1.2           & $30.6 \pm 11.2$ & 30(2) \\
HD\,13246\,B      & 02 07 32.2 & $-$59 40 23 & 286,$-$55 & $22.6 \pm 0.5$  & 1  & K5Ve  & 7.5 & $0.8 \pm 0.5$               &               & \\
HD\,17925         & 02 52 32.1 & $-$12 46 11 & 192,$-$58 & $96.6 \pm 0.4$  & 1  & K1V   & 4.2 & 0.9           & $34.4 \pm 20.4$ & 41(3), 40(4), 40(5)\\
HD\,19668         & 03 09 42.3 & $-$09 34 47 & 191,$-$53 & $26.7 \pm 1.1$  & 1  & G0V   & 6.7 & 1.1           & $31.8 \pm 1.1$ & 70(2)\\
1E\,0307.4+1424   & 03 10 12.5 & $+$14 36 03 & 166,$-$36 & $6.3 \pm 4.2$   & 1  & G6V   & 8.8 & 1.1               & & \\
HD\,283798        & 04 41 55.2 & $+$26 58 49 & 174,$-$13 & $9.0 \pm 1.3$   & 2  & G2V   & 8.0 & 1.2           & $16.1 \pm 2.1$ & 11(6)\\
HD\,33802\,B      & 05 12 17.6 & $-$11.51 58 & 213,$-$27 & $14.1 \pm 0.2$  & 1  & G5Ve  & 7.8 & 1.1               & & \\
HD\,41842         & 06 06 16.6 & $-$27 54 21 & 234,$-$22 & $31.2 \pm 1.0$  & 1  & K1V   & 6.8 & 0.8           & $60.6 \pm 11.1$ & \\
HD\,47787         & 06 39 11.6 & $-$26 34 19 & 236,$-$14 & $19.4 \pm 1.6$  & 2  & K1V   & 6.5 & 1.1           & $16.5 \pm 6.5$ & \\
GJ\,426 A           & 11 21 49.3 & $+$18 11 24 & 232,$+$67 & $31.7 \pm 1.5$  & 2  & G5V & 6.0 & $ 1.2 \pm 0.3$               & & \\
GJ\,466           & 12 25 58.6 & $+$08 03 44 & 284,$+$70 & $26.7 \pm 2.3$  & 2  & M0V   & 7.2 & $ 0.6 \pm 0.3$               & & \\
HD\,109138        & 12 33 29.8 & $-$75 23 11 & 301,$-$13 & $16.7 \pm 1.0$  & 1  & K0V   & 7.8 & 0.9           & $23.9 \pm 10.0$ & \\
HD\,119022        & 13 43 08.7 & $-$69 07 39 & 308,$-$07 & $8.1 \pm 0.8$   & 2  & G5IV  & 5.8 & 1.2               & & \\
HD\,137727        & 15 28 44.0 & $-$31 17 38 & 338,$+$21 & $8.9 \pm 1.9$   & 2  & K0    & 6.9 & 1.5           & $8.2 \pm 0.6$ & \\
HD\,138009        & 15 30 26.2 & $-$32 18 13 & 338,$+$19 & $9.8 \pm 1.9$   & 2  & G6V   & 6.9 & 1.4           & $10.8 \pm 4.4$ & \\
HD\,139084        & 15 38 57.5 & $-$57 42 27 & 324,$-$02 & $26.0 \pm 1.1$  & 1  & K0V   & 5.9 & 1.0           & $20.0 \pm 5.4$ & \\
HD\,197890        & 20 47 45.0 & $-$36 35 41 & 006,$-$38 & $19.2 \pm 1.8$  & 1  & K3Ve  & 6.8 & 1.0           & $24.1 \pm 9.3$ & \\
HD\,202917        & 21 20 50.0 & $-$53 02 03 & 344,$-$44 & $23.3 \pm 1.0$  & 1  & G7V   & 6.9 & 1.0           & $39.6 \pm 9.9$ & 15(7), 30(2), 10(8)\\
FG\,Aqr           & 22 17 19.0 & $-$08 48 12 & 052,$-$49 & $96.0 \pm 6.0$  & 3  & M4V   & 8.2 & 0.2               & & \\
HD\,217352        & 22 59 59.4 & $+$05 09 34 & 079,$-$48 & $5.3 \pm 0.7$   & 1  & K1III & 4.4 & 5.0           &  & Post-MS(1) \\
HD\,222259        & 23 39 39.5 & $-$69 11 45 & 312,$-$47 & $21.9 \pm 0.8$  & 2  & G5V   & 6.7 & 1.1           & $20.5 \pm 6.5$ & \\
\hline\hline
\newline
\end{tabular}\label{table_targets}
\begin{flushleft}
References: (1) Kumar et al. (2011), (2) Chen et al. (2014), (3) Lachaume  et al. (1999): lower mass limit, (4) Lafreniere et al. (2007): lower mass limit, (5) Bryden et al. (2006): lower mass limit, (6) Palla \& Stahler (2002), (7) Maldonado et al. (2012), (8) Wright et al. (2004). \\ 
\end{flushleft}   
\end{table}
\end{landscape}
\clearpage

\begin{table*}

\caption{The astrometric calibration of NACO for all observing epochs, whose data are presented here.\newline The pixel scale $PS$ and the detector position angle $DPA$ (defined from N over E to S) are listed with their uncertainties.\newline}

\begin{tabular}{ccc}
\hline
Epoch   & $PS$               & $DPA$ \\
        & [mas/pixel]        & [$^{\circ}$] \\
\hline
2009/09 & $13.234 \pm 0.018$ & $0.42 \pm 0.10$ \\
2009/12 & $13.231 \pm 0.025$ & $0.25 \pm 0.14$ \\
2010/02 & $13.232 \pm 0.018$ & $0.37 \pm 0.10$ \\
2010/03 & $13.239 \pm 0.022$ & $0.63 \pm 0.15$ \\
2010/05 & $13.231 \pm 0.020$ & $0.67 \pm 0.13$ \\
2010/10 & $13.227 \pm 0.029$ & $0.57 \pm 0.13$ \\
2011/03 & $13.231 \pm 0.022$ & $0.69 \pm 0.15$ \\
\hline\hline
\end{tabular}\label{table_astrocal}
\end{table*}
\clearpage

\begin{table*}
\centering
\caption{Magnitude difference $\Delta K_{s}$ vs. limiting angular separation between host star and companion, based on the dynamical ranges of our deep NACO $K_{s}$-band imaging data, achieved after the subtraction of the point spread function of the bright target stars.\newline}
\begin{tabular}{cccccc}
\hline
Target & &\multicolumn{4}{c}{$\Delta K_{s}$}\\
 & Sep [arcsec] & 0.2 & 0.5 & 0.8 & 1.8 \\
\hline
HD\,225213 & & 5.1 & 7.5 & 10.7 & 12.3 \\
HD\,105    & &6.6 & 8.7 & 11.5 & 12.3 \\
HD\,6665   & &6.8 & 8.7 & 11.8 & 12.8 \\
BD+17\,232  & & 7.1 & 9.6 & 10.7 & 10.7 \\
HD\,8558    & &6.6 & 8.8 & 10.9 & 11.3 \\
HD\,9054    & &6.9 & 9.0 & 11.6 & - \\
HD\,13183    & &7.0 & 8.9 & 10.6 & 11.1 \\
HD\,13246\,A  & &6.3 & 8.6 & 10.0 & 11.1 \\
HD\,13246\,B  & &6.3 & 8.3 & 9.8 & 10.0 \\
HD\,17925    & &5.2 & 7.3 & 10.0 & 12.0 \\
HD\,19668    & &7.1 & 9.3 & 12.0 & 12.5 \\
1E\,0307.4+1424 & & 7.2 & 9.8 & 11.2 & 11.3 \\
HD\,283798    & &6.3 & 8.1 & 9.6 & 9.7 \\
HD\,33802\,B  & &7.4 & 10.0 & 12.1 & 12.2 \\
HD\,41842     & &7.0 & 9.7 & 12.4 & 12.9 \\
HD\,47787     & &6.8 & 9.9 & 11.1 & 10.9 \\
GJ\,426       & &6.3 & 9.0 & 11.4 & 12.0 \\
GJ\,466       & &6.6 & 9.7 & 11.6 & 11.6 \\
HD\,109138    & &6.5 & 8.1 & 9.1 & 9.5 \\
HD\,119022    & &5.2 & 9.0 & 11.4 & 12.6 \\
HD\,137727    & &6.6 & 9.5 & 11.1 & 11.5 \\
HD\,138009    & &6.2 & 8.9 & 10.7 & 10.8 \\
HD\,139084    & &6.8 & 8.5 & 11.2 & 12.2  \\
HD\,197890    & &5.9 & 9.3 & 11.8 & 12.2   \\
HD\,202917    & &7.2 & 9.5 & 12.0 & 12.6 \\
FG\,Aqr       & &6.5 & 9.5 & 10.8 & -  \\
HD\,217352    & &5.1 & 7.5 & 10.6 & 12.1 \\
HD\,222259    & &6.3 & 8.4 & 10.5 & 11.0 \\
\hline
\hline
\end{tabular}
\end{table*}
\clearpage

\begin{table*}
\centering
\caption{This table lists for each observed target the number of observing epochs ($N_{ep}$), the number of exposures per jitter-position (NDIT), the number of jitter-positions (NINT) for each epoch, the number of detected companion-candidates $N_{cc}$, detected in the NACO S13 FOV i.e. with separations from 0.2 to 8 arcsec, as well as the detection limits, achieved with NACO in the K$_{\rm s}$-band in the background limited region around each target. The last six columns summarizes the minimum-masses of detectable sub-stellar companions of all targets in the background limited region, as well as the minimal projected separations to the host star, beyond which all stellar companions (mass $>75\,M_{\rm Jup}$) are detectable, for assumed ages of the targets of 10, 50, and 100\,Myr. The achieved dynamic ranges for all targets are shown in Fig.\,\ref{figure_dranges}.\newline}

\begin{tabular}{ccccccccccc}
\hline
Target & $N_{ep}$ & NDIT,NINT & $N_{cc}$ & $K_{s_{Limit}}$ & \multicolumn{3}{c}{Detection Limit [$M_{Jup}$]} & \multicolumn{3}{c}{$sep_{min}$ for 75\,$M_{Jup}$ [AU]}\\
& & 1st Ep $|$ 2nd Ep &  & [mag] & 10\,Myr & 50\,Myr & 100\,Myr & 10\,Myr & 50\,Myr & 100\,Myr\\
\hline
HD\,225213      & 1 & 126,42                      & 0  & 16.7 & 2  & 4  & 5  & $<$1 & $<$1 & $<$1 \\
HD\,105         & 1 & 126,42                      & 0  & 18.2 & 3  & 7  & 9  & 4   & 6   & 6 \\
HD\,6665        & 1 & 126,56                      & 0  & 18.3 & 8  & 13 & 25 & 76  & 145 & 171 \\
BD+17\,232      & 1 & 126,41                      & 0  & 17.4 & 5  & 10 & 12 & 7   & 10  & 11 \\
HD\,8558        & 1 & 126,42                      & 0  & 18.1 & 3  & 7  & 10 & 4   & 6   & 7 \\
HD\,9054        & 2 & 126,42 $|$ 126,30     & 1  & 18.7 & 2  & 6  & 8  & 3   & 4   & 4 \\
HD\,13183       & 1 & 126,77                      & 0  & 18.0 & 3  & 8  & 11 & 4   & 6   & 7 \\
HD\,13246\,A    & 1 & 126,42                      & 0  & 17.2 & 4  & 9  & 11 & 4   & 6   & 7 \\
HD\,13246\,B    & 1 & 126,42                      & 0  & 17.5 & 4  & 8  & 11 & 3   & 4   & 5 \\
HD\,17925       & 1 & 126,33                      & 0  & 16.0 & 2  & 6  & 8  & 1   & 2   & 2 \\
HD\,19668       & 2 & 126,52 $|$ 126,30     & 1  & 19.0 & 2  & 6  & 8  & 3   & 4   & 4 \\
1E\,0307.4+1424 & 1 & 126,42                      & 0  & 20.0 & 4  & 9  & 11 & 16  & 19  & 24 \\
HD\,283798      & 2 & 126,41 $|$ 126,39     & 3  & 17.7 & 7  & 12 & 18 & 12  & 19  & 21 \\
HD\,33802\,B    & 1 & 126,42                & 0  & 20.0 & 3  & 6  & 8  & 6   & 7   & 9 \\
HD\,41842       & 1 & 126,42                & 0  & 19.7 & 2  & 5  & 7  & 2   & 3   & 3 \\
HD\,47787       & 2 & 126,41 $|$ 126,9      & 0  & 17.6 & 4  & 9  & 11 & 5   & 7   & 8 \\
GJ\,426         & 2 & 126,42 $|$ 126,42     & 0  & 19.2 & 2  & 5  & 7  & 2   & 3   & 4 \\
GJ\,466         & 1 & 126,42                & 0  & 18.9 & 2  & 6  & 8  & 1   & 1   & 3 \\
HD\,109138      & 2 & 126,32 $|$ 126,42     & 1  & 17.1 & 5  & 11 & 12 & 4   & 7   & 8 \\
HD\,119022      & 2 & 126,42 $|$ 126,42     & 1  & 18.3 & 6  & 11 & 12 & 37  & 53  & 59 \\
HD\,137727      & 1 & 126,15                & 0  & 18.2 & 5  & 11 & 12 & 18  & 24  & 26 \\
HD\,138009      & 1 & 126,41                & 0  & 17.5 & 6  & 12 & 17 & 16  & 22  & 28 \\
HD\,139084      & 2 & 126,42 $|$ 126,21     & 12 & 18.0 & 3  & 7  & 10 & 4   & 5   & 6 \\
HD\,197890      & 1 & 126,42                & 0  & 19.0 & 3  & 6  & 9  & 5   & 7   & 10 \\
HD\,202917      & 1 & 126,42                & 0  & 19.3 & 2  & 6  & 8  & 3   & 5   & 6 \\
FG\,Aqr         & 2 & 126,42 $|$ 126,31     & 1  & 19.0 & 2  & 3  & 5  & $<$1 & $<$1 & $<$1 \\
HD\,217352      & 1 & 126,42                      & 0  & 16.6 & 13 & 28 & 40 & 150 & 180 & 228 \\
HD\,222259      & 1 & 126,42                      & 0  & 17.6 & 4  & 8  & 11 & 4   & 5   & 6 \\
\hline\hline

\end{tabular}\label{table_results}

\end{table*}
\clearpage

\begin{table*}
\centering
\caption{Measured angular separations ($sep$) and position angles ($PA$) of stellar binary- and triple-systems for all observing epochs.\newline}

\begin{tabular}{ccccc}
\hline
Target     & Components & $JD$         & $sep$             & $PA$ \\
           &           & $-$2455000 & [arcsec]             & [$^{\circ}$]\\
\hline
BD+17\,232 & A/B & 103.5 & $1.662 \pm 0.003$ & $203.56 \pm 0.11$\\
\\
HD 283798     & A/B & 173.5 & $1.640 \pm 0.003$ & $301.52 \pm 0.14$\\
           &     & 644.5 & $1.639 \pm 0.003$ & $301.52 \pm 0.16$\\
\\
HD 47787     & A/B & 173.5 & $2.110 \pm 0.004$ & $201.54 \pm 0.14$\\
           &     & 645.5 & $2.122 \pm 0.004$ & $201.57 \pm 0.16$\\
\\
GJ\,426    & A/B & 251.5 & $5.089 \pm 0.007$ & $314.60 \pm 0.10$\\
           &     & 263.5 & $5.087 \pm 0.009$ & $314.63 \pm 0.15$\\
\\
GJ\,466    & A/B & 263.5 & $1.931 \pm 0.003$ & $107.98 \pm 0.15$\\
\\
HD 119022     & A/B & 250.5 & $0.210 \pm 0.001$ & $32.56 \pm 0.21$\\
           &     & 645.5 & $0.215 \pm 0.001$ & $33.22 \pm 0.20$\\
\\
HD 137727     & A/B & 251.5 & $2.198 \pm 0.003$ & $184.27 \pm 0.10$\\
\\
HD 138009     & A/B & 251.5 & $1.600 \pm 0.002$ & $22.81 \pm 0.11$\\
\\
FG\,Aqr    & A/B & 173.5 & $7.895 \pm 0.015$ & $213.15 \pm 0.14$\\
           &     & 498.5 & $7.907 \pm 0.017$ & $213.40 \pm 0.13$\\
\\
FG\,Aqr    & B/C & 173.5 & $0.947 \pm 0.002$ & $318.52 \pm 0.17$\\
           &     & 498.5 & $0.924 \pm 0.003$ & $320.09 \pm 0.18$\\
\\
HD 222259     & A/B & 104.5 & $5.334 \pm 0.008$ & $347.61 \pm 0.10$\\
\hline\hline
\end{tabular}\label{table_binastro}
\end{table*}
\clearpage

\begin{table*}
\centering
\caption{The astrometry of all detected companion-candidates for both observing epochs.\newline}

\begin{tabular}{cc|ccc|ccc}
\hline
       &           &  \multicolumn{3}{c|}{\large{First epoch}} & \multicolumn{3}{c}{\large{Second epoch}} \\
Target & Candidate & $JD$ & $sep$ & $PA$ & $JD$ & $sep$ & $PA$\\
       &           & $-$2455000 & [arcsec] & [$^{\circ}$] & $-$2455000 & [arcsec] & [$^{\circ}$]\\
\hline
HD\,9054   &              cc01                  & 104.5                 & $3.832 \pm 0.006$ & $104.13 \pm 0.11$ & 498.5 & $3.728 \pm 0.008$ & $103.84 \pm 0.13$\\
\multicolumn{2}{c|}{} & \multicolumn{3}{c|}{} & \multicolumn{3}{c}{}\\
HD\,19668  &              cc01                  & 103.5                 & $5.573 \pm 0.008$ & $150.45 \pm 0.10$ & 498.5 & $5.424 \pm 0.012$ & $150.53 \pm 0.13$\\
\multicolumn{2}{c|}{} & \multicolumn{3}{c|}{} & \multicolumn{3}{c}{}\\
HD\,283798 &              cc01                  & 173.5                 & $2.546 \pm 0.005$ & $131.87 \pm 0.14$ & 644.5 & $2.529 \pm 0.009$ & $131.38 \pm 0.24$\\
           &              cc02                  &                       & $7.324 \pm 0.014$ & $62.34 \pm 0.14$  &       & $7.346 \pm 0.012$ & $62.76 \pm 0.15$\\
           &              cc03                  &                       & $7.768 \pm 0.015$ & $76.28 \pm 0.14$  &       & $7.779 \pm 0.014$ & $76.16 \pm 0.15$\\
\multicolumn{2}{c|}{} & \multicolumn{3}{c|}{} & \multicolumn{3}{c}{}\\
HD\,109138 &              cc01                  & 173.5                 & $2.912 \pm 0.006$ & $13.90 \pm 0.16$  & 645.5 & $2.938 \pm 0.005$ & $16.10 \pm 0.15$\\
\multicolumn{2}{c|}{} & \multicolumn{3}{c|}{} & \multicolumn{3}{c}{}\\
HD\,119022 &              cc01                  & 250.5                 & $5.748 \pm 0.008$ & $9.50 \pm 0.10$   & 645.5 & $5.764 \pm 0.010$ & $9.82 \pm 0.15$\\
\multicolumn{2}{c|}{} & \multicolumn{3}{c|}{} & \multicolumn{3}{c}{}\\
HD\,139084 &              cc01                  & 103.5                 & $6.820 \pm 0.009$ & 70.89$\pm$0.10    & 645.5 & $6.917 \pm 0.012$ & $69.87 \pm 0.15$\\
           &              cc02                  &                       & $6.519 \pm 0.010$ & 59.82$\pm$0.11    &       & $6.630 \pm 0.011$ & $58.87 \pm 0.15$\\
           &              cc03                  &                       & $4.348 \pm 0.006$ & 79.92$\pm$0.10    &       & $4.426 \pm 0.007$ & $77.99 \pm 0.15$\\
           &              cc04                  &                       & $4.670 \pm 0.007$ & 63.78$\pm$0.10    &       & $4.786 \pm 0.008$ & $62.51 \pm 0.15$\\
           &              cc05                  &                       & $3.049 \pm 0.005$ & 351.57$\pm$0.11   &       & $3.217 \pm 0.005$ & $352.86 \pm 0.15$\\
           &              cc06                  &                       & $5.523 \pm 0.008$ & 336.35$\pm$0.10   &       & $5.665 \pm 0.009$ & $3370.52 \pm 0.15$\\
           &              cc07                  &                       & $7.547 \pm 0.011$ & 298.40$\pm$0.10   &       & $7.592 \pm 0.013$ & $299.80 \pm 0.15$\\
           &              cc08                  &                       & $3.660 \pm 0.006$ & 257.97$\pm$0.11   &       & $3.589 \pm 0.006$ & $260.64 \pm 0.15$\\
           &              cc09                  &                       & $2.224 \pm 0.003$ & 196.59$\pm$0.11   &       & $2.051 \pm 0.004$ & $196.88 \pm 0.15$\\
           &              cc10                  &                       & $4.612 \pm 0.007$ & 211.23$\pm$0.10   &       & $4.443 \pm 0.007$ & $212.07 \pm 0.15$\\
           &              cc11                  &                       & $4.462 \pm 0.008$ & 203.96$\pm$0.11   &       & $4.230 \pm 0.007$ & $204.51 \pm 0.15$\\
           &              cc12                  &                       & $6.673 \pm 0.010$ & 184.29$\pm$0.10   &       & $6.497 \pm 0.011$ & $184.20 \pm 0.15$\\
\multicolumn{2}{c|}{} & \multicolumn{3}{c|}{} & \multicolumn{3}{c}{}\\
FG\,Aqr    &              cc01                  & 173.5                 & $2.347 \pm 0.005$ & 118.69$\pm$0.16   & 498.5 & $2.609 \pm 0.007$ & $110.02 \pm 0.22$\\
\hline\hline
\end{tabular}\label{table_ccastro}
\end{table*}
\clearpage

\begin{figure*}[h!]
\centering
\caption{Typical examples of the curves displaying the dynamical ranges of our deep NACO K$_{\rm s}$-band imaging data magnitude difference $\triangle{}K_{s}$, plotted versus angular (\textit{bottom x-axis}) and projected (\textit{top x-axis}) separation (in AU) from all targets, achieved after the substraction of the point spread function of the bright target stars. For the $\triangle{}K_{s}$ values of selected angular separations of all targets see Table 3. \newline}\label{figure_dranges}

\resizebox{\hsize}{!}{\includegraphics{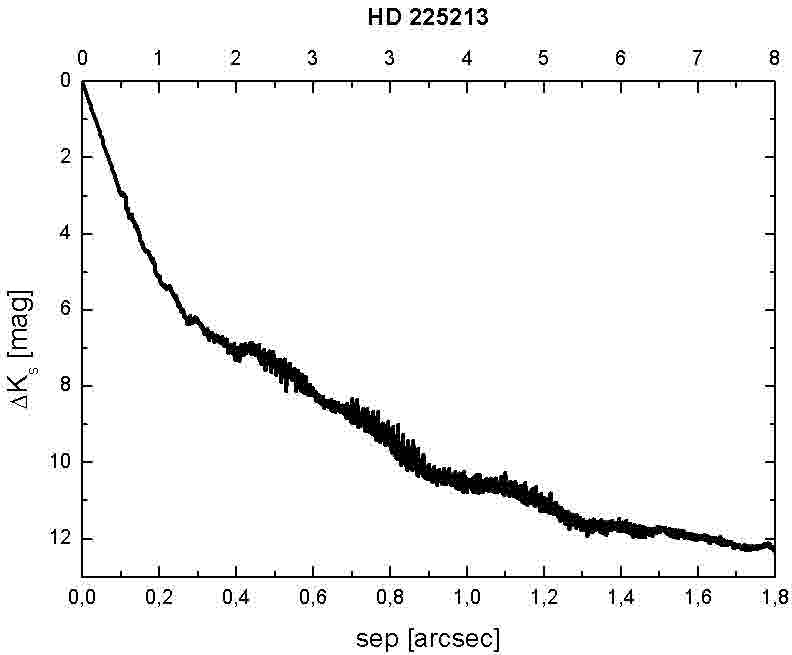}\includegraphics{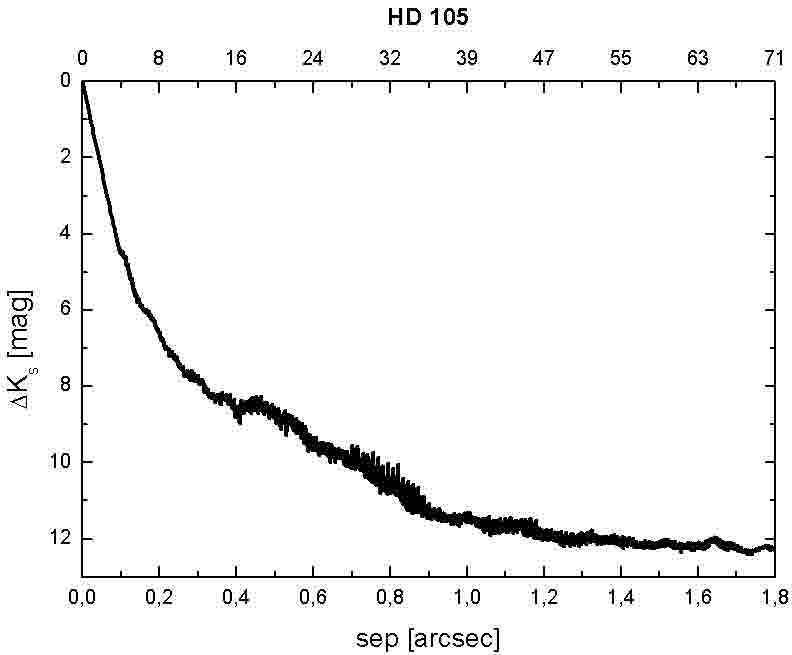}}
\resizebox{\hsize}{!}{\includegraphics{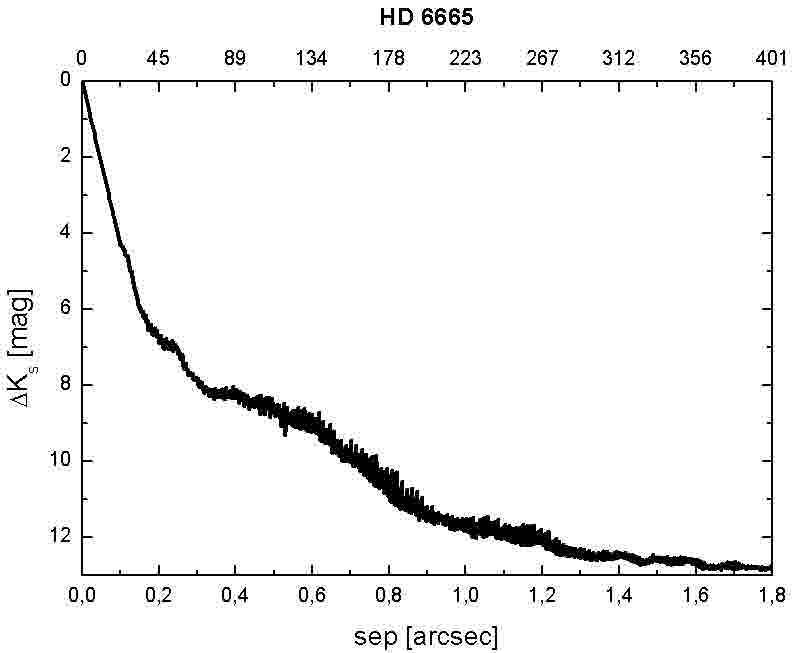}\includegraphics{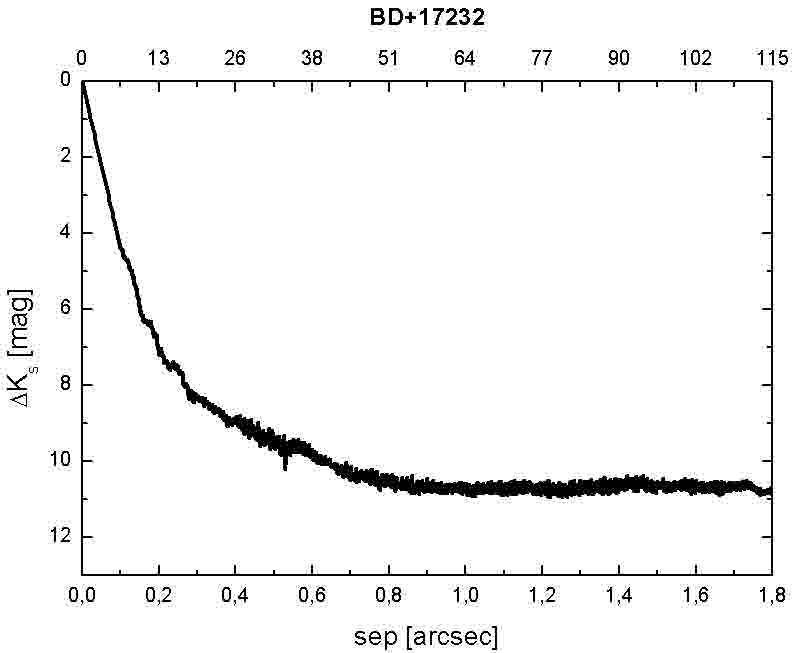}}
\end{figure*}
\clearpage

\begin{figure*}
\centering
\caption{The angular separations (\textit{left}) and the position angles (\textit{right}) vs. Julian date for stellar binary and or triple system observed during two different epochs (c.f. Tab.\,\ref{table_binastro}). The wobbled lines illustrate the expected changes in both quantities in the case that the companion is a non-moving background object. The dotted straight lines reproduce the error bar of the first epoch observation and mark the range of a co-moving companion without significant orbital motion. The cones, marked by two solid lines, show the expected maximum change in both parameters for a companion on a gravitationally bound orbit (i.e. in a circular edge-on orbit for separation or a circular pole-on orbit for PA). All cases presented here are compatible with physically bound binary configurations, FG\,Aqr is a hierarchic triple system.\newline} \label{figure_binastro}

\resizebox{\hsize}{!}{\includegraphics{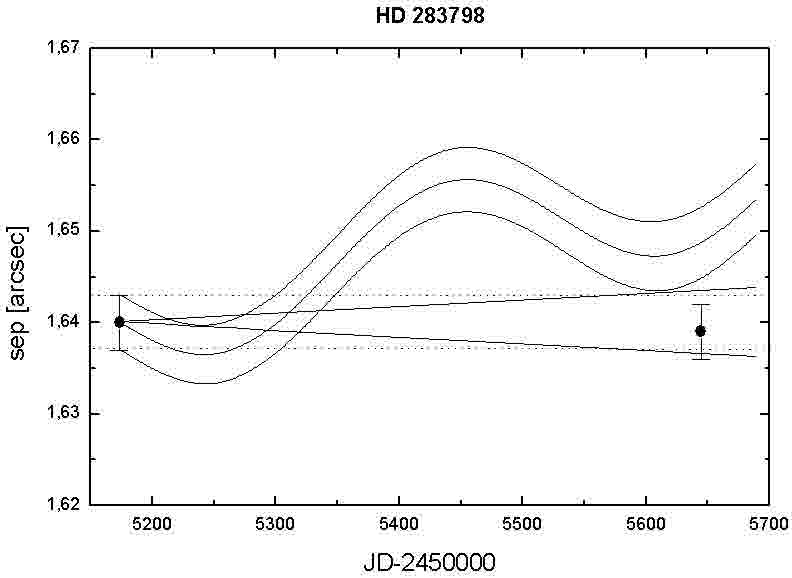}\includegraphics{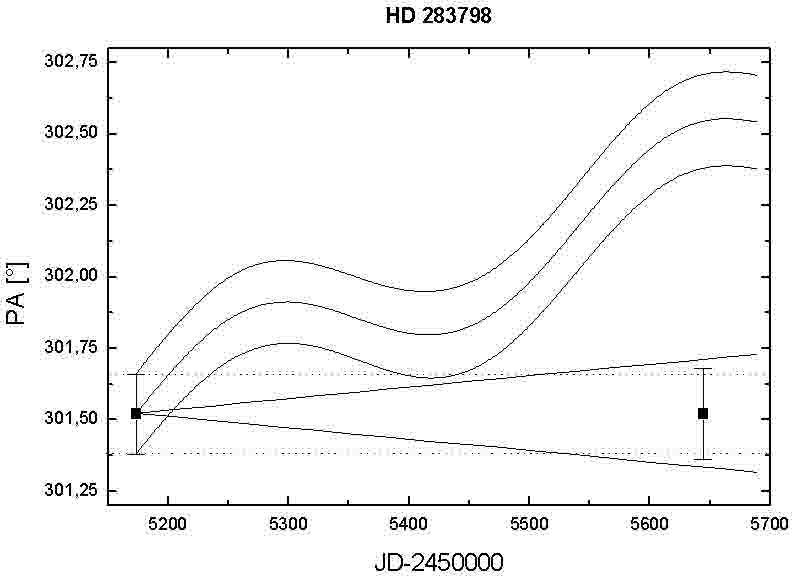}}
\resizebox{\hsize}{!}{\includegraphics{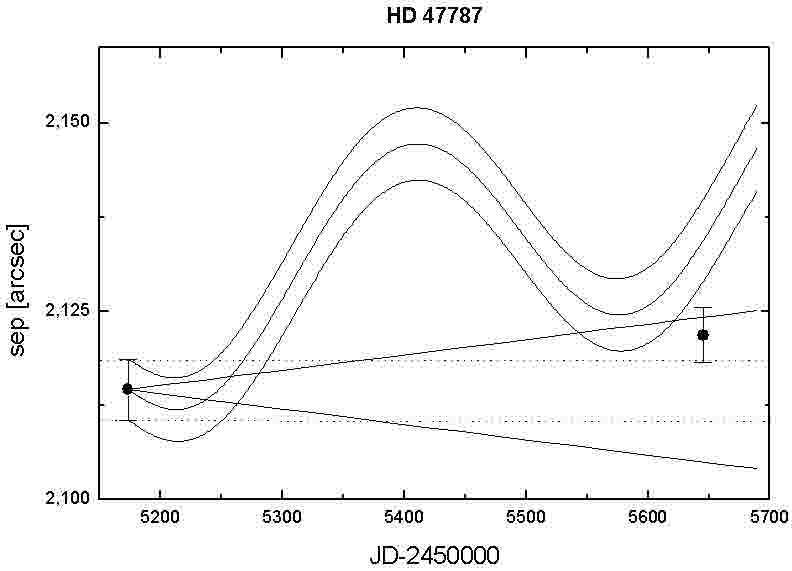}\includegraphics{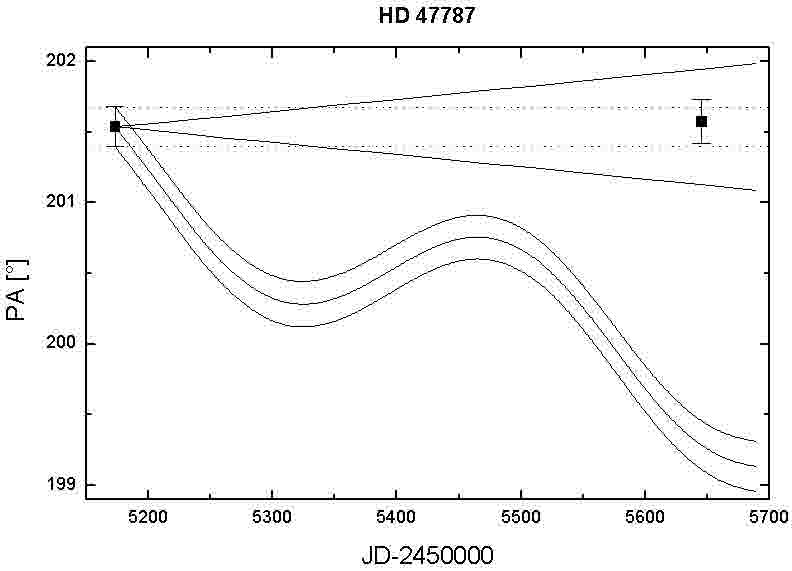}}
\resizebox{\hsize}{!}{\includegraphics{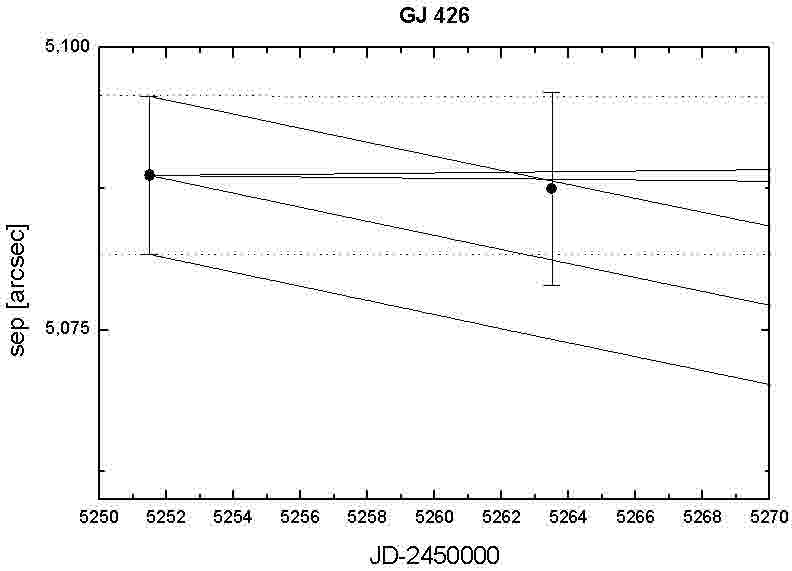}\includegraphics{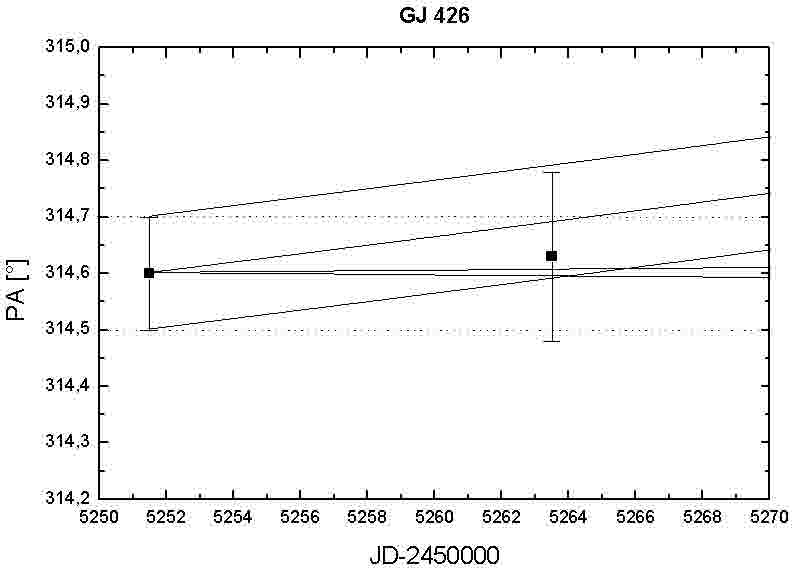}}
\end{figure*}
\clearpage

\begin{figure*}
\resizebox{\hsize}{!}{\includegraphics{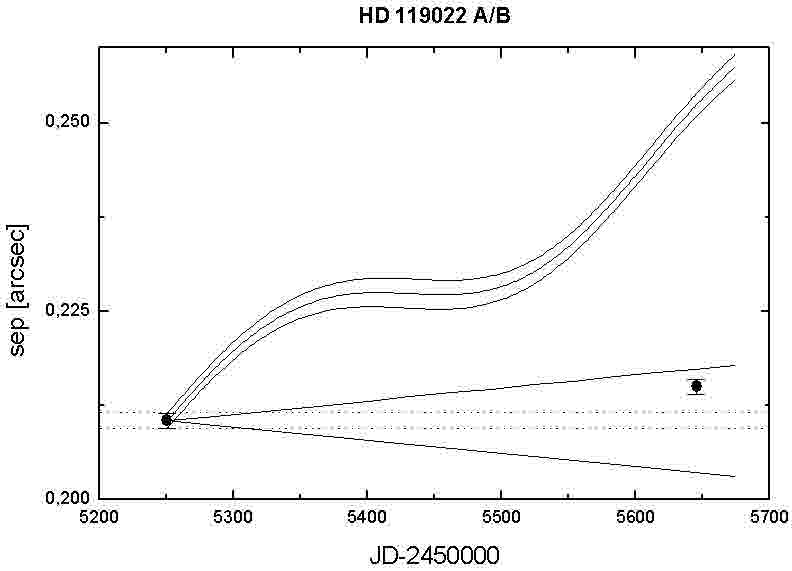}\includegraphics{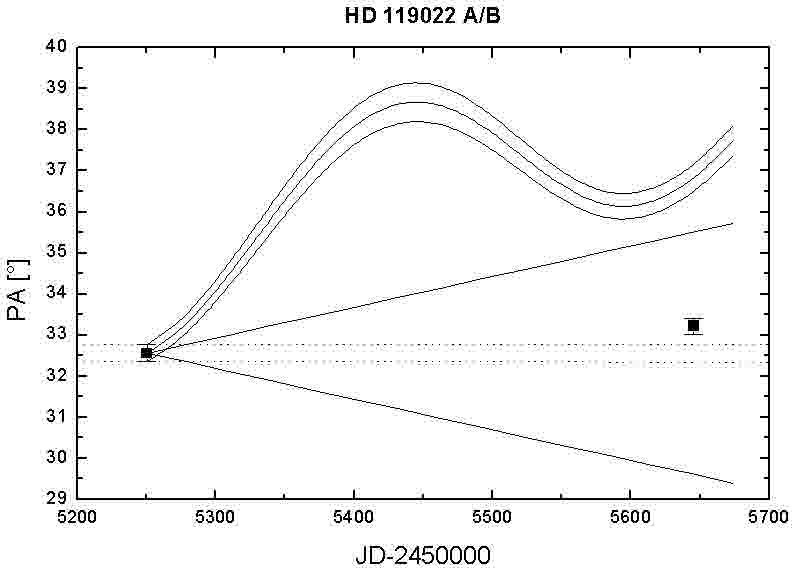}}
\resizebox{\hsize}{!}{\includegraphics{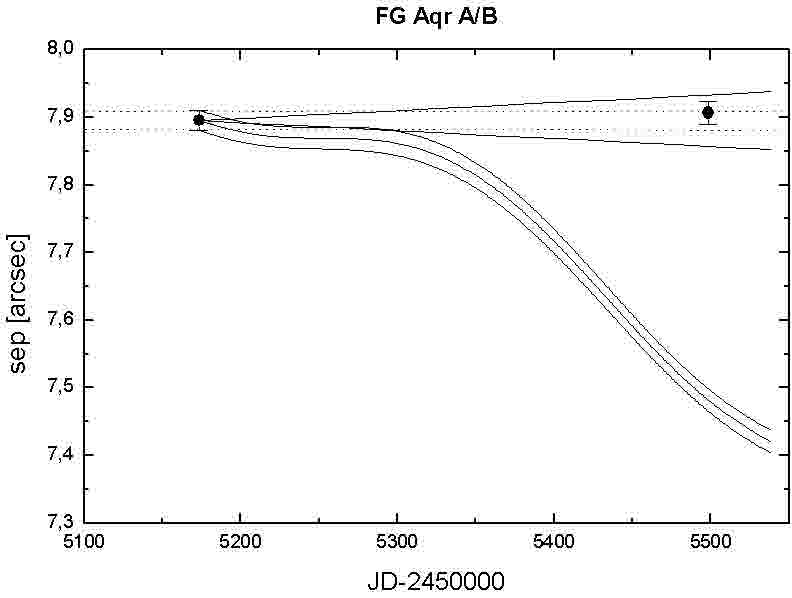}\includegraphics{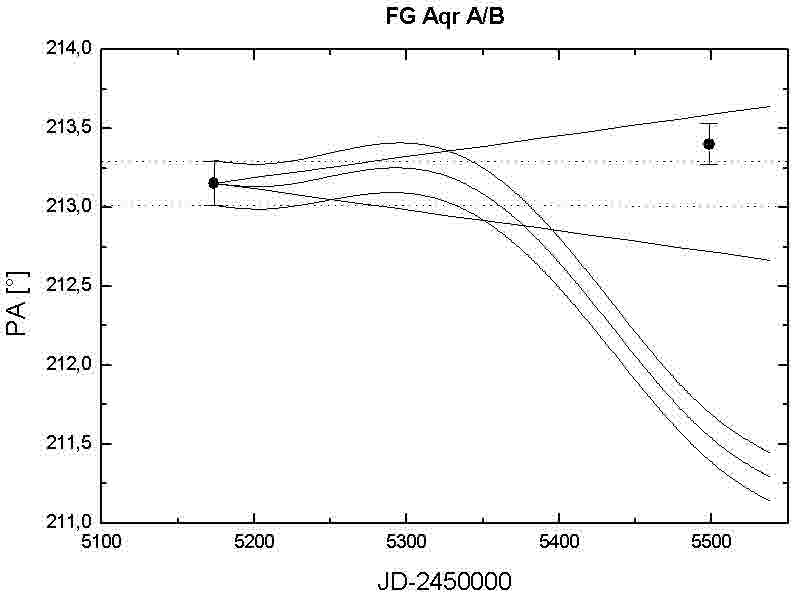}}
\resizebox{\hsize}{!}{\includegraphics{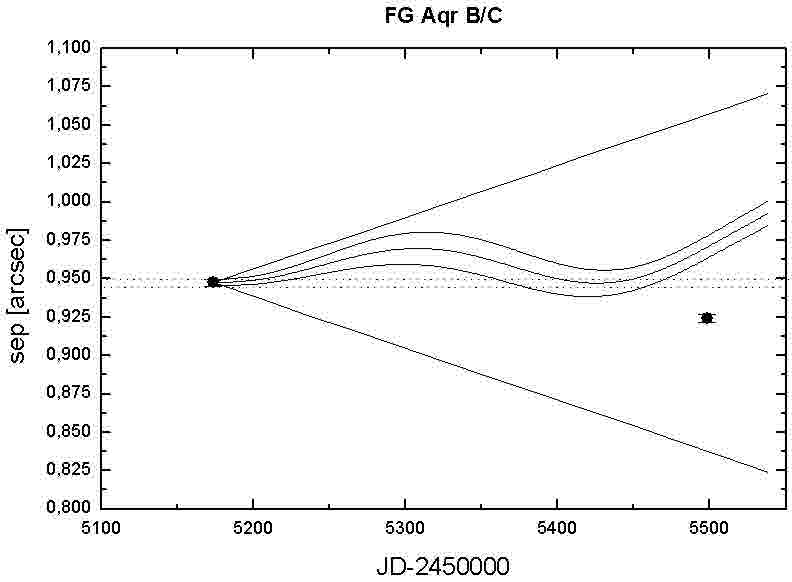}\includegraphics{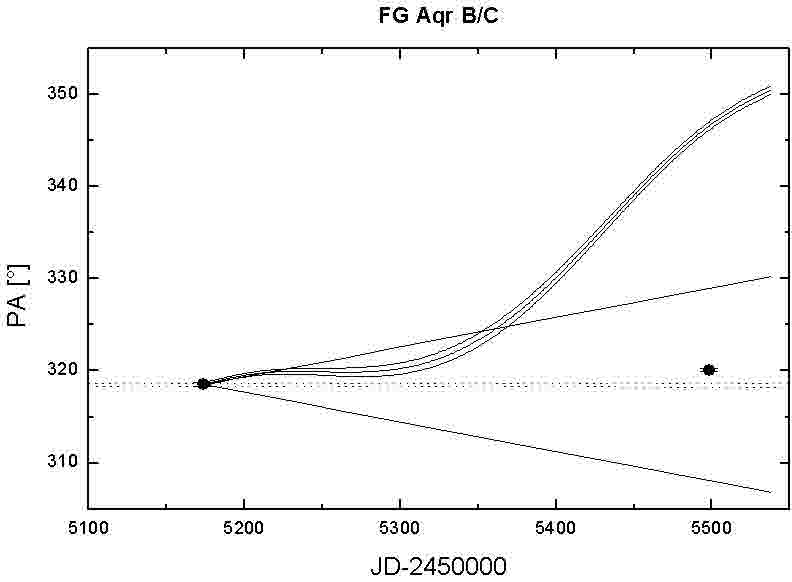}}
\end{figure*}
\clearpage

\begin{figure*}
\centering
\caption{The angular separations (\textit{left}) and the position angles (\textit{right}) vs. Julian date for 20 sub-stellar companion-candidates found in this study (c.f. Tab.\,\ref{table_ccastro}), in the same presentation as in Fig.\,\ref{figure_binastro}. All candidates turn out to be background objects, there is no indication for any object gravitationally bound to its host star.\newline}\label{figure_ccastro}

\resizebox{\hsize}{!}{\includegraphics{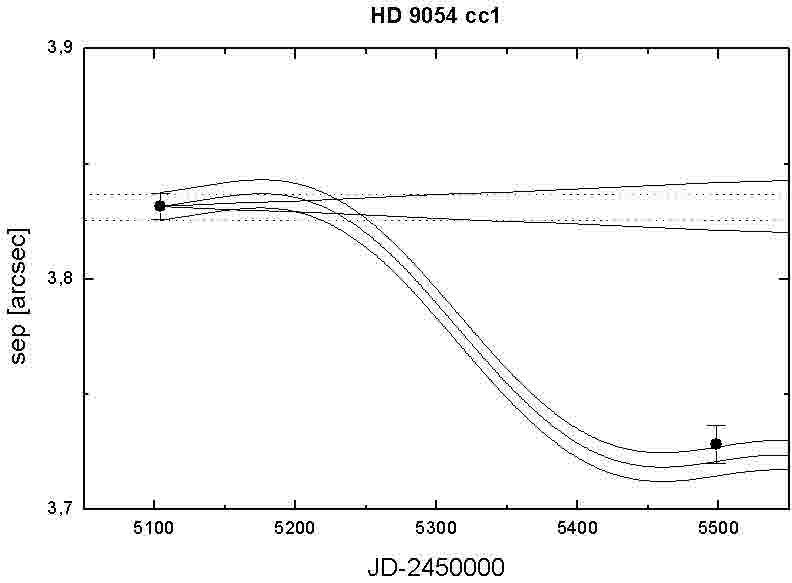}\includegraphics{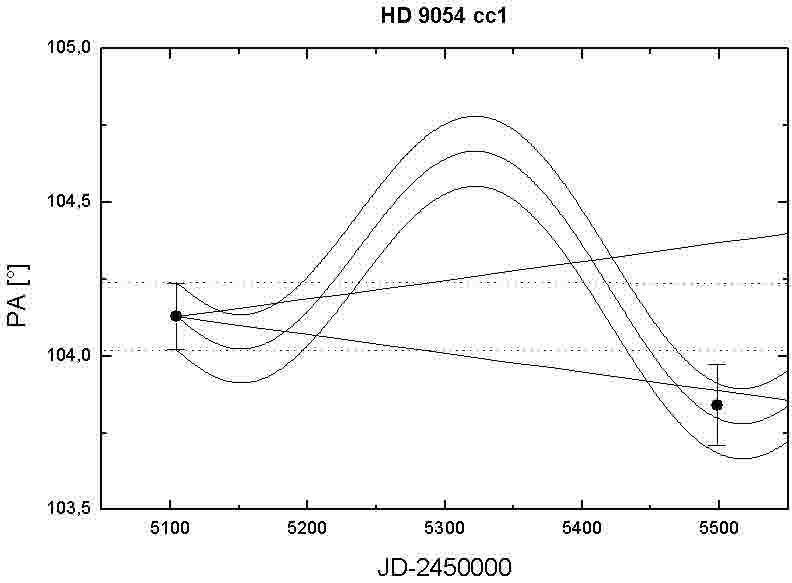}}
\resizebox{\hsize}{!}{\includegraphics{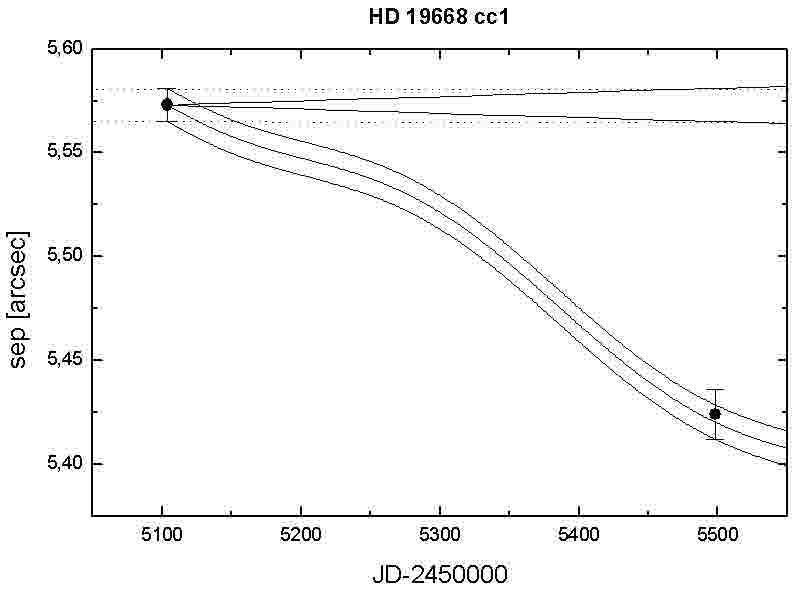}\includegraphics{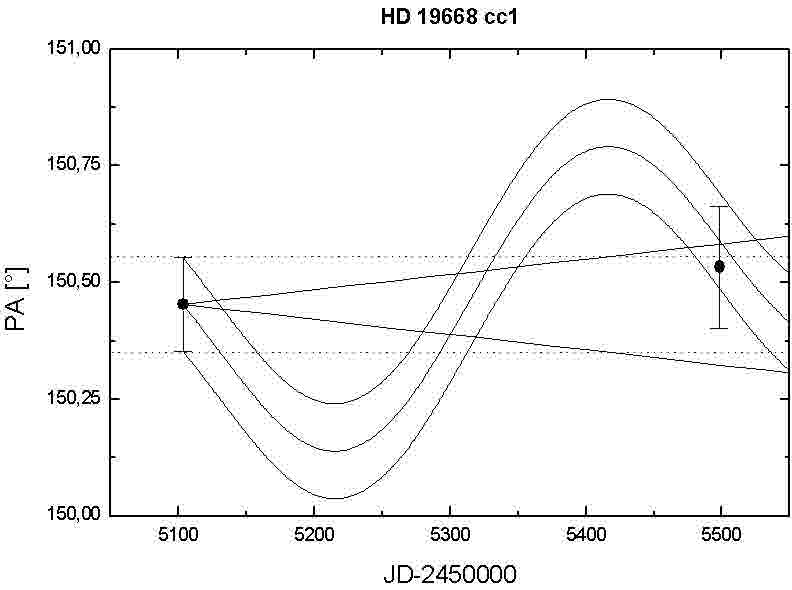}}
\resizebox{\hsize}{!}{\includegraphics{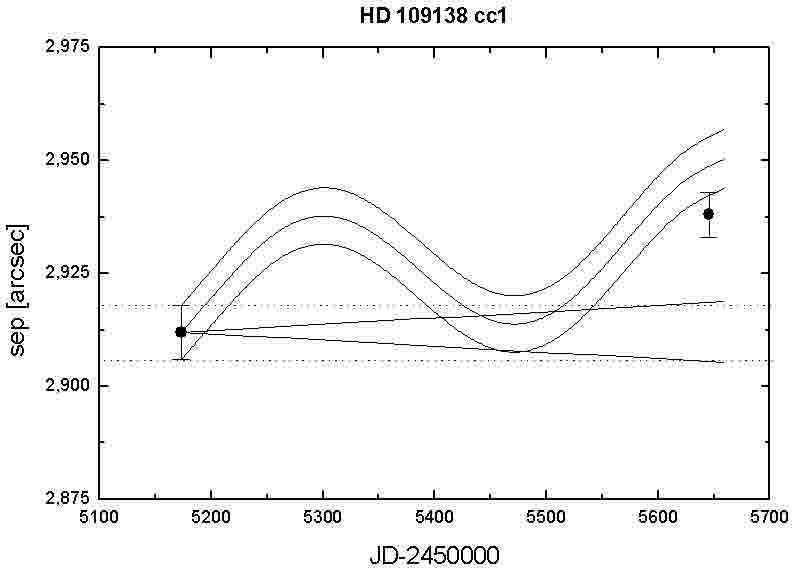}\includegraphics{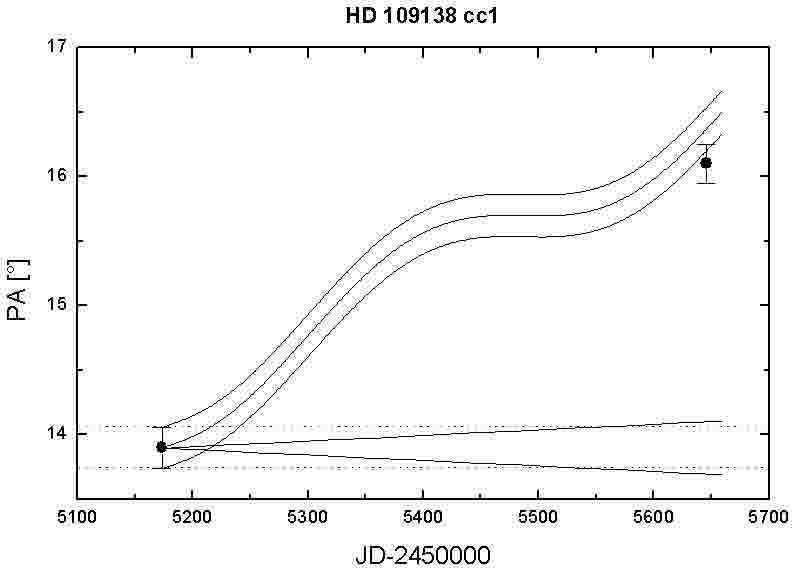}}
\end{figure*}
\clearpage

\begin{figure*}
\resizebox{\hsize}{!}{\includegraphics{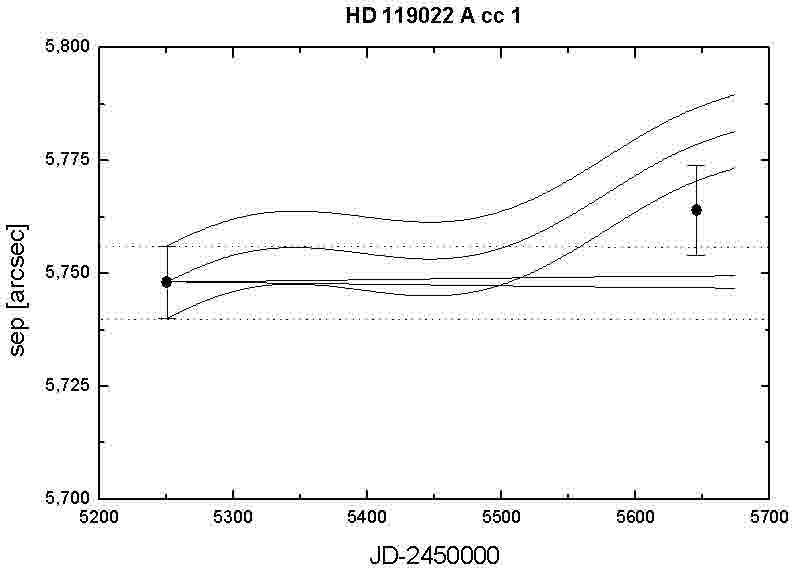}\includegraphics[angle=0]{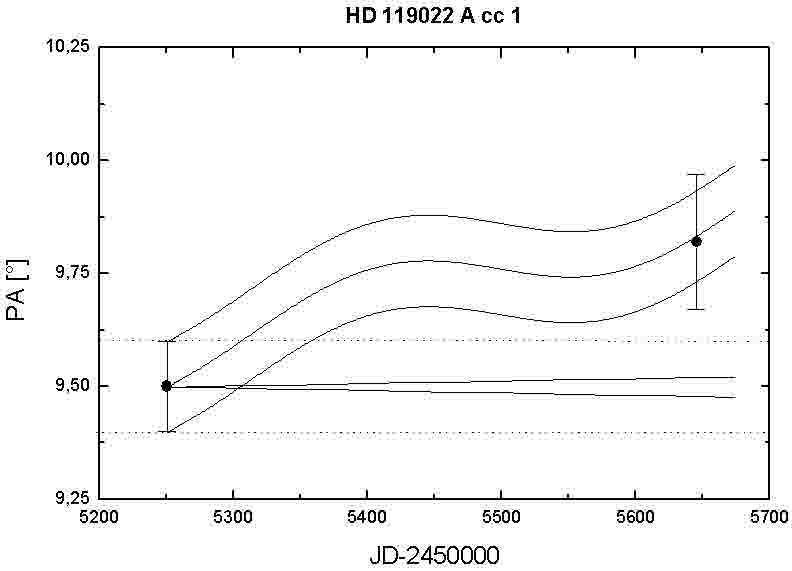}}
\resizebox{\hsize}{!}{\includegraphics{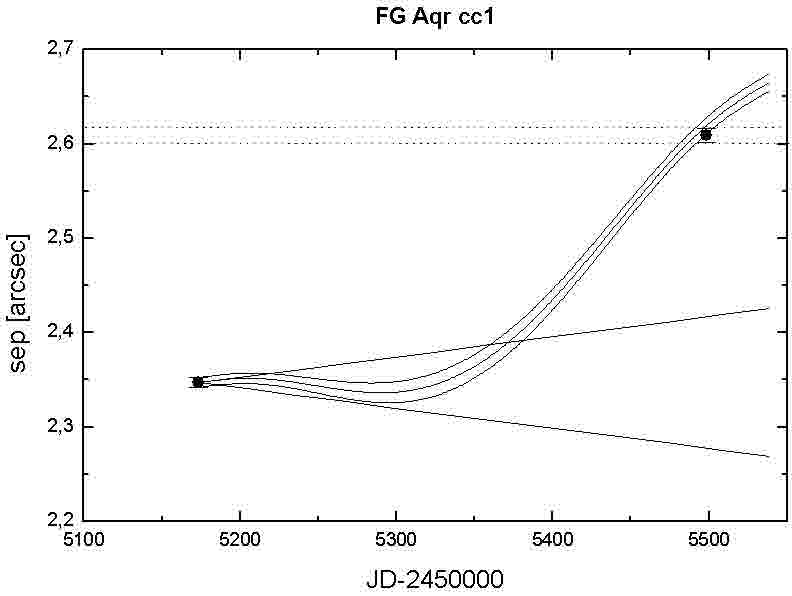}\includegraphics[angle=0]{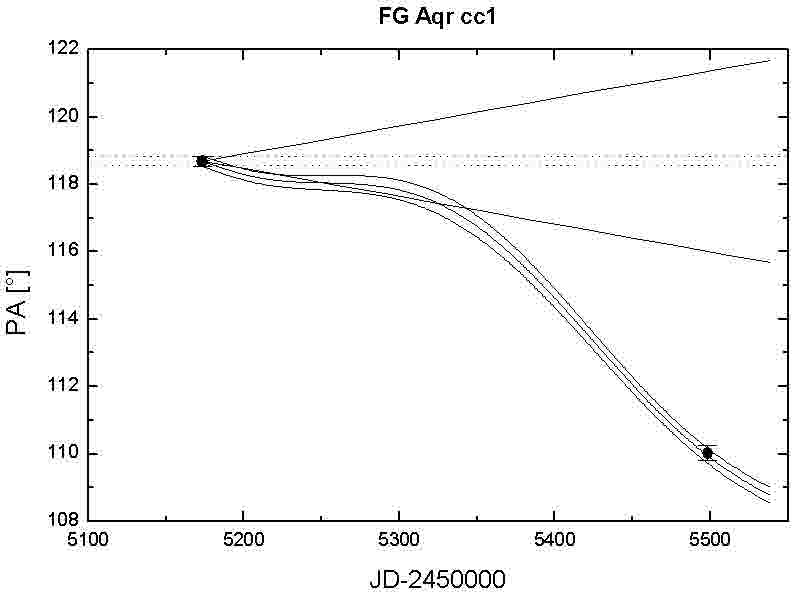}}
\end{figure*}
\clearpage

\begin{figure*}
\resizebox{\hsize}{!}{\includegraphics{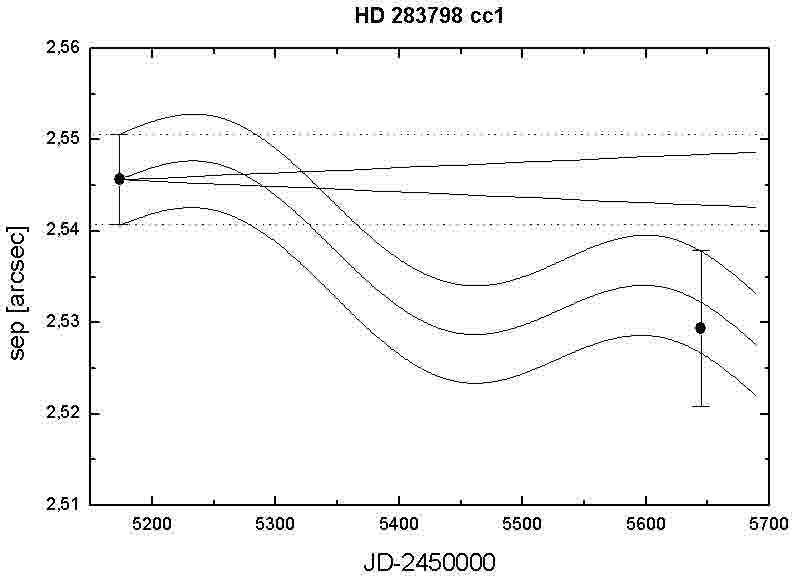}\includegraphics[angle=0]{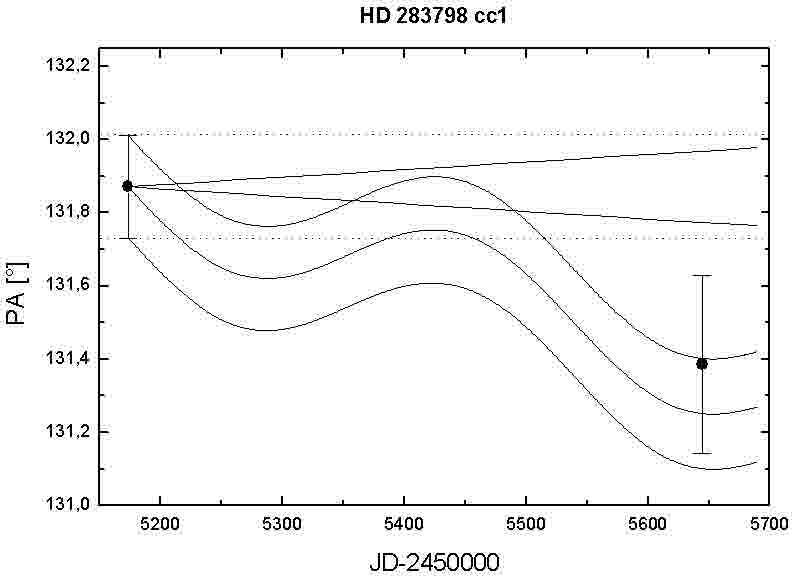}}
\resizebox{\hsize}{!}{\includegraphics{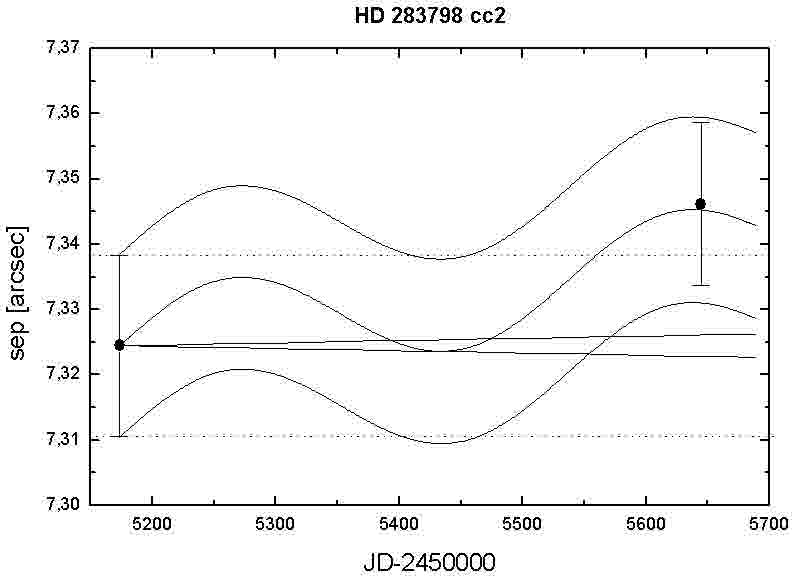}\includegraphics[angle=0]{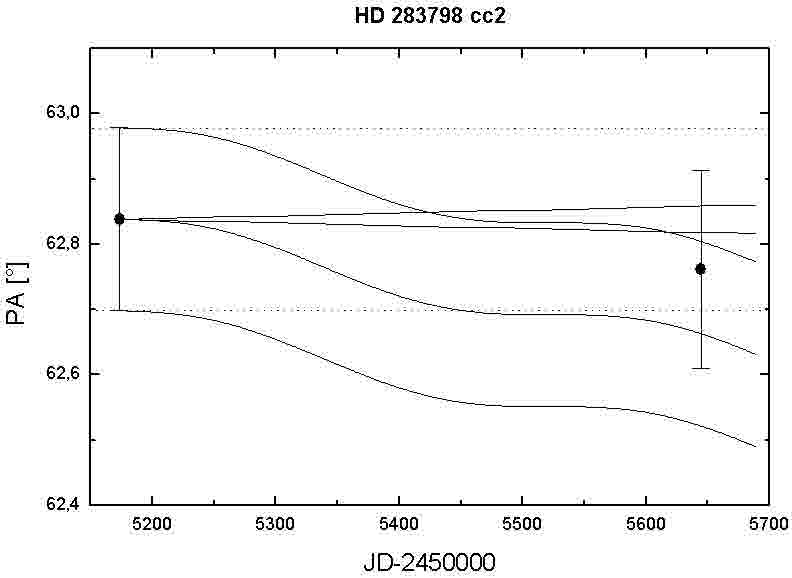}}
\resizebox{\hsize}{!}{\includegraphics{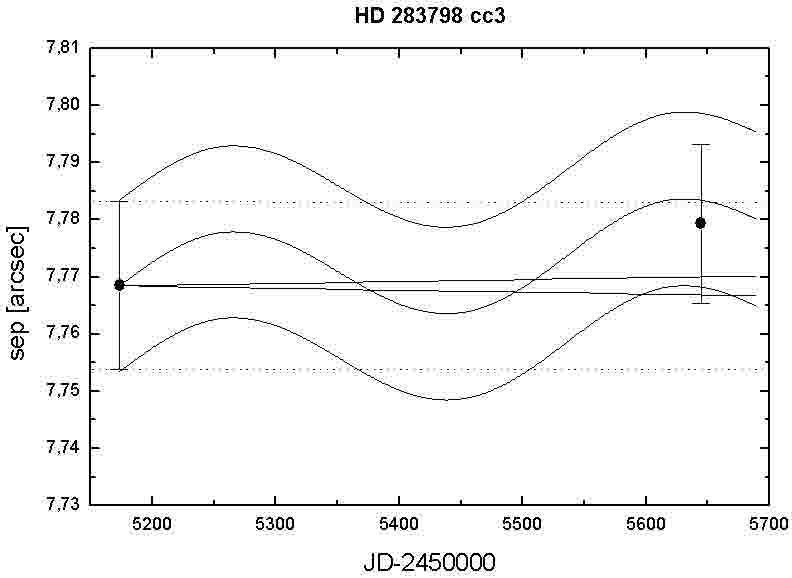}\includegraphics[angle=0]{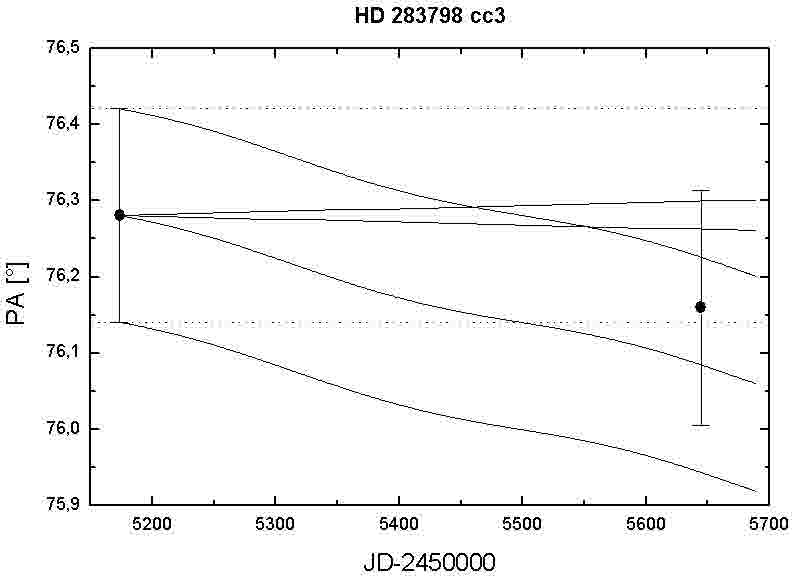}}
\end{figure*}
\clearpage

\begin{figure*}
\resizebox{\hsize}{!}{\includegraphics{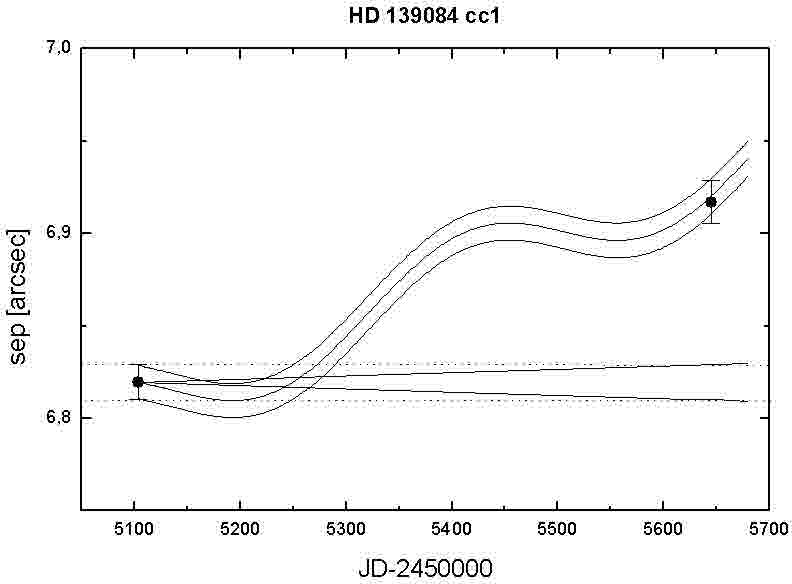}\includegraphics[angle=0]{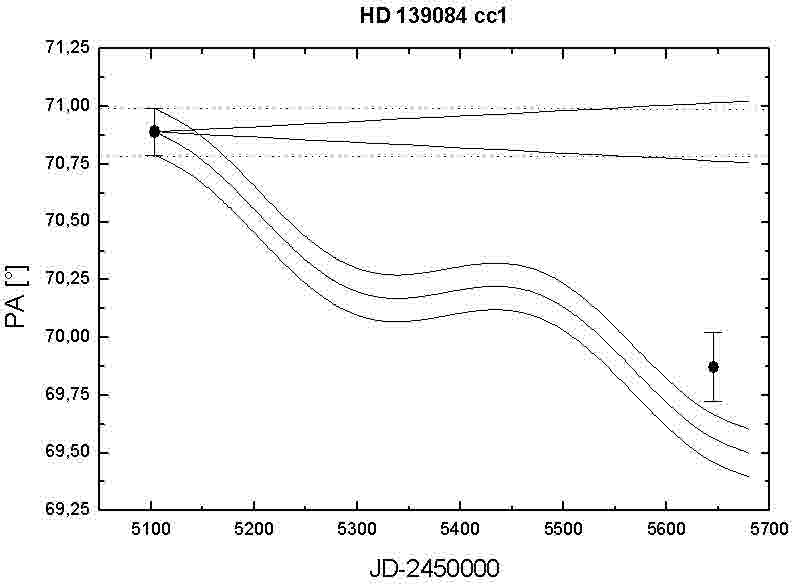}}
\resizebox{\hsize}{!}{\includegraphics{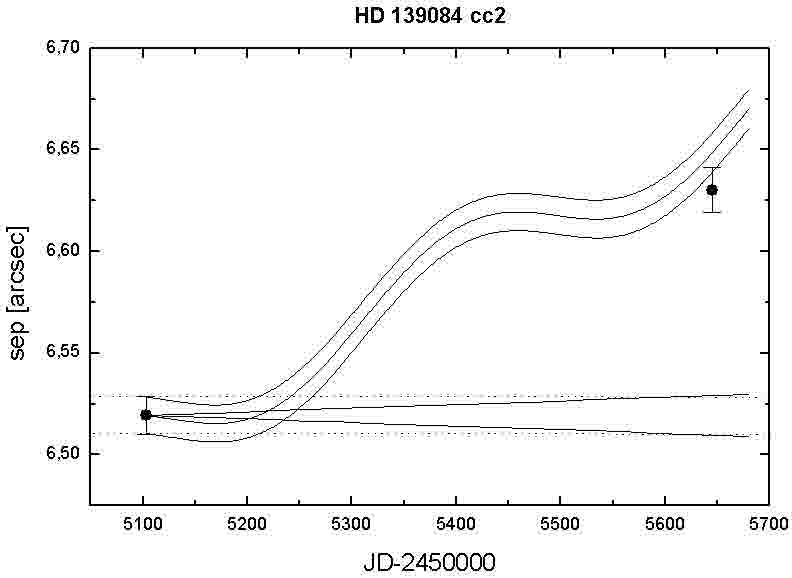}\includegraphics[angle=0]{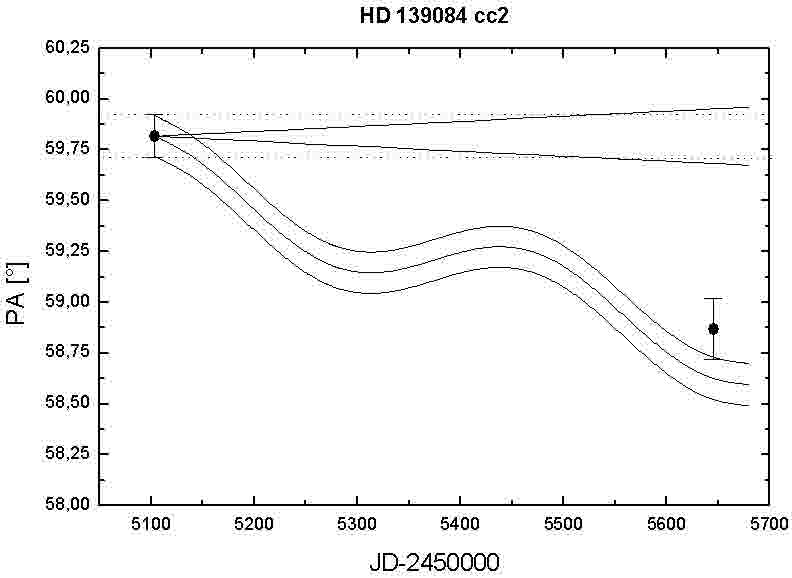}}
\resizebox{\hsize}{!}{\includegraphics{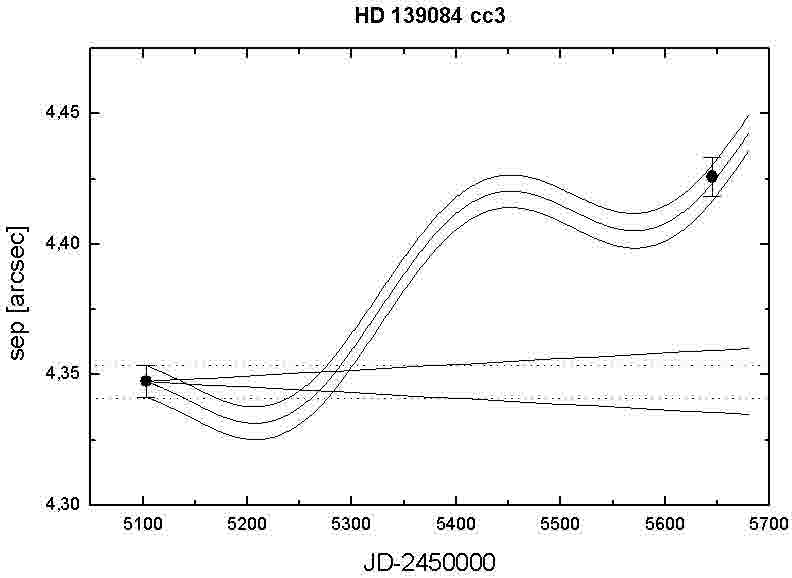}\includegraphics[angle=0]{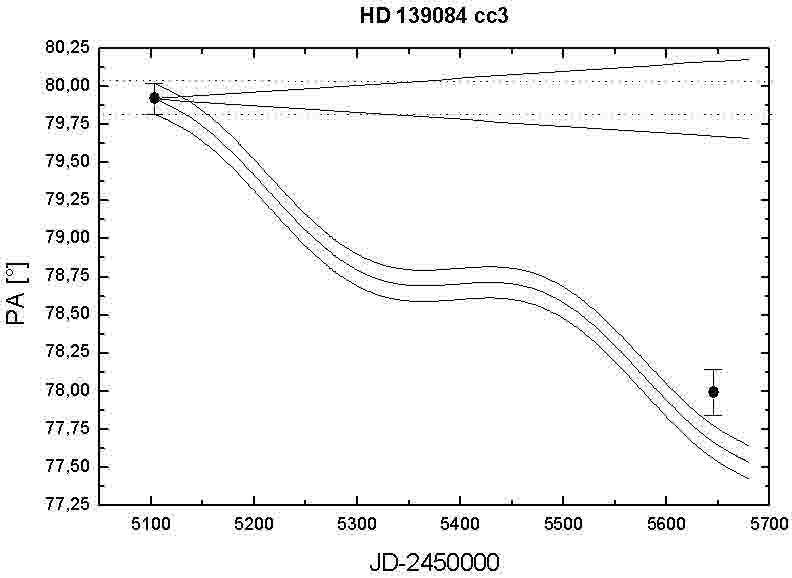}}
\end{figure*}
\clearpage

\begin{figure*}
\resizebox{\hsize}{!}{\includegraphics{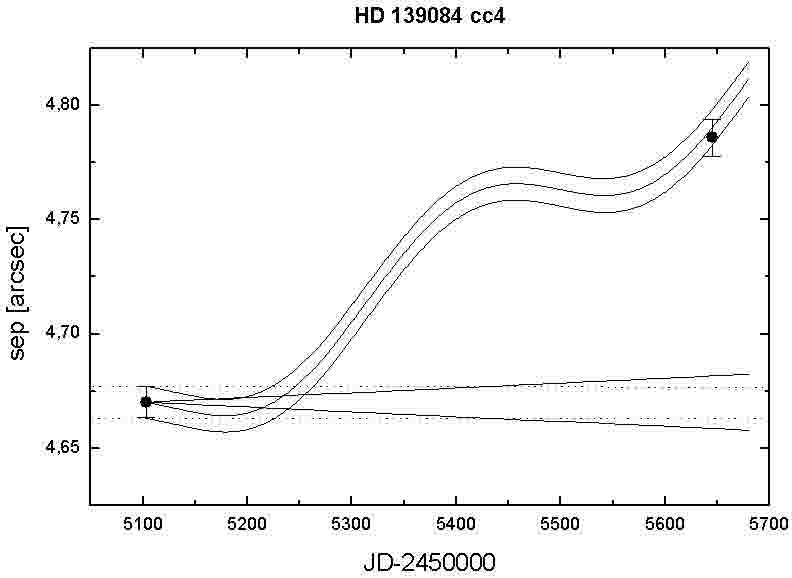}\includegraphics[angle=0]{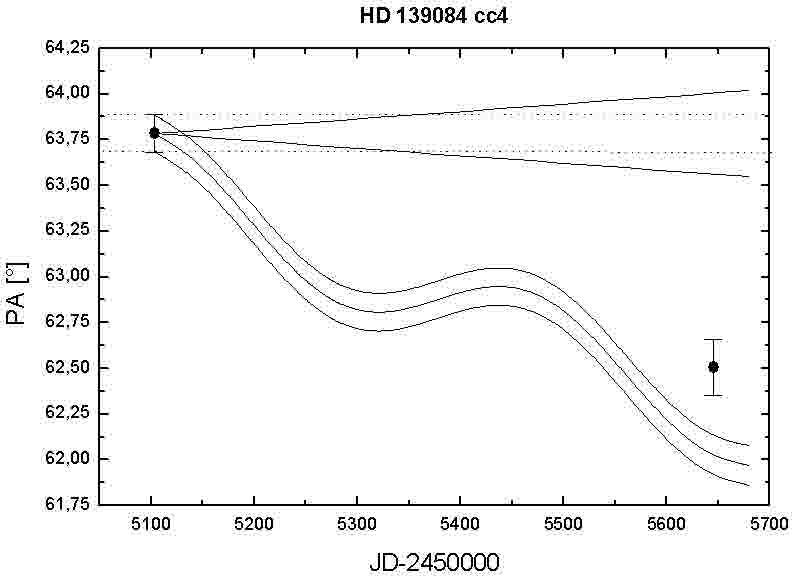}}
\resizebox{\hsize}{!}{\includegraphics{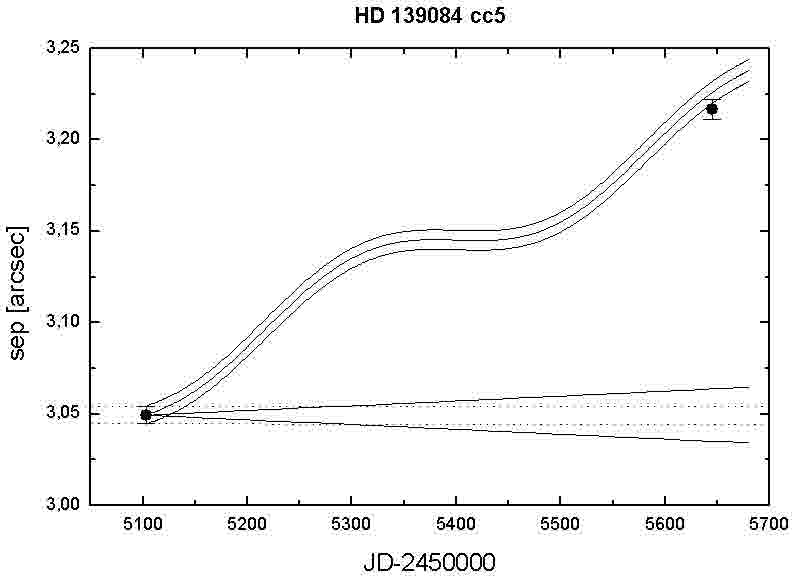}\includegraphics[angle=0]{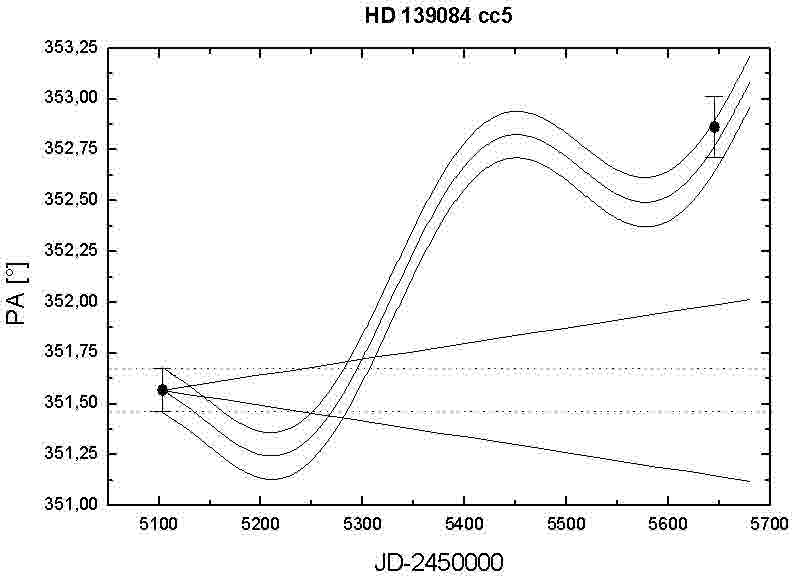}}
\resizebox{\hsize}{!}{\includegraphics{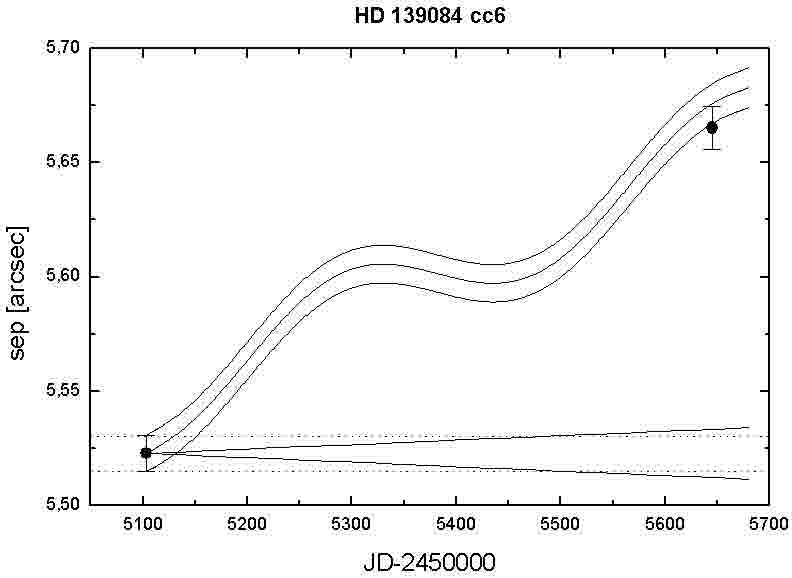}\includegraphics[angle=0]{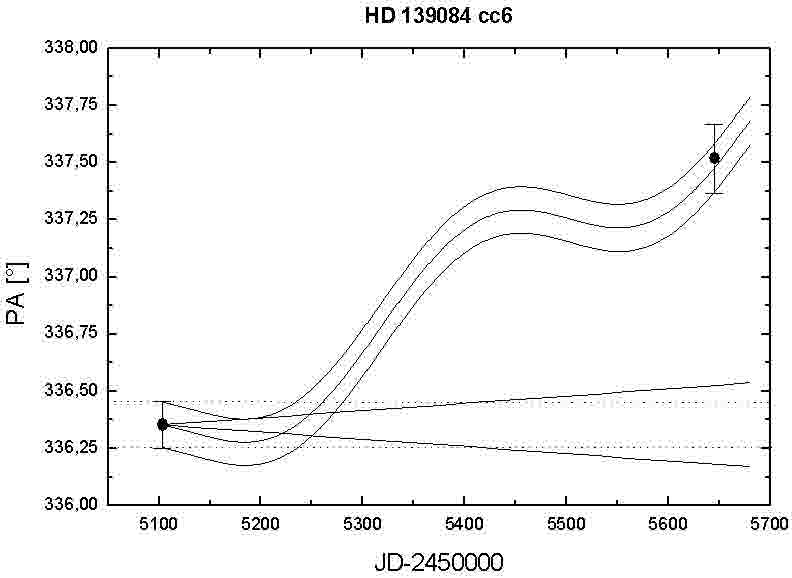}}
\end{figure*}
\clearpage

\begin{figure*}
\resizebox{\hsize}{!}{\includegraphics{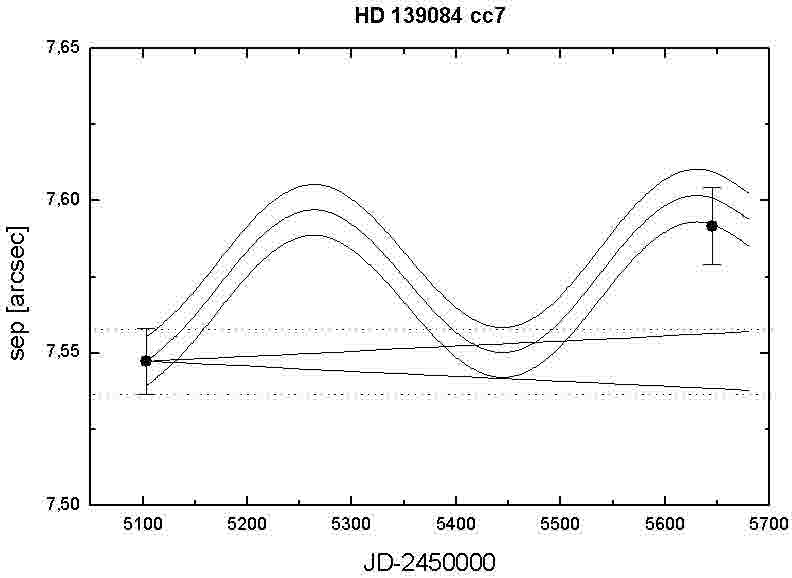}\includegraphics[angle=0]{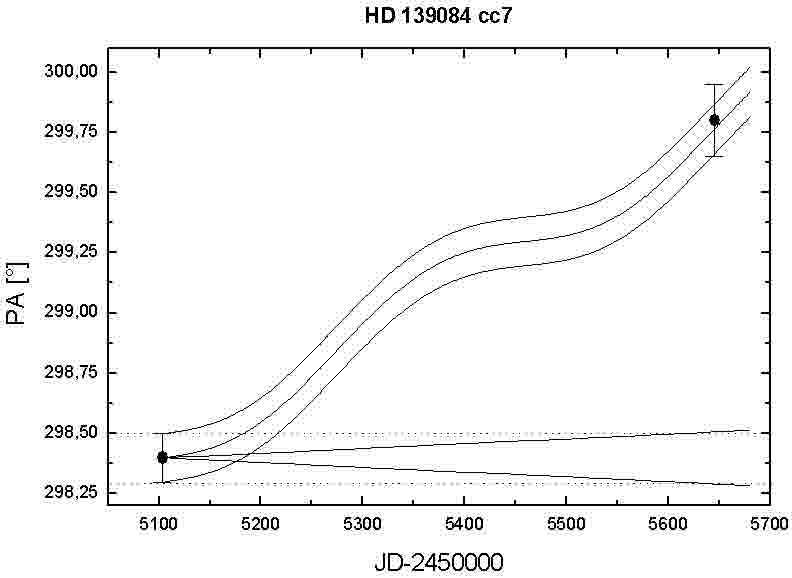}}
\resizebox{\hsize}{!}{\includegraphics{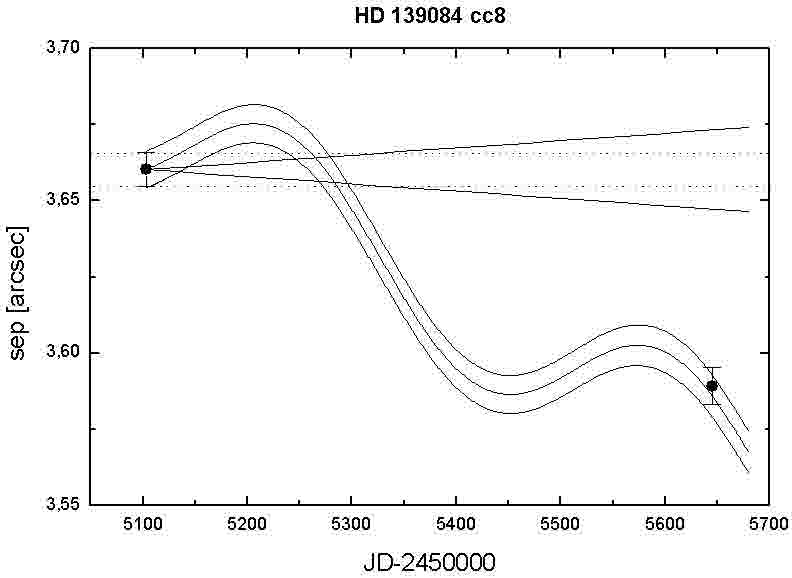}\includegraphics[angle=0]{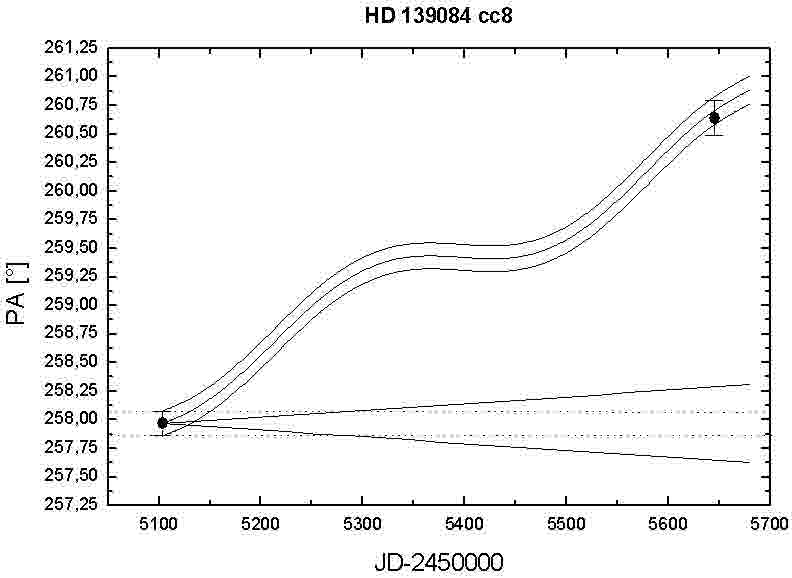}}
\resizebox{\hsize}{!}{\includegraphics{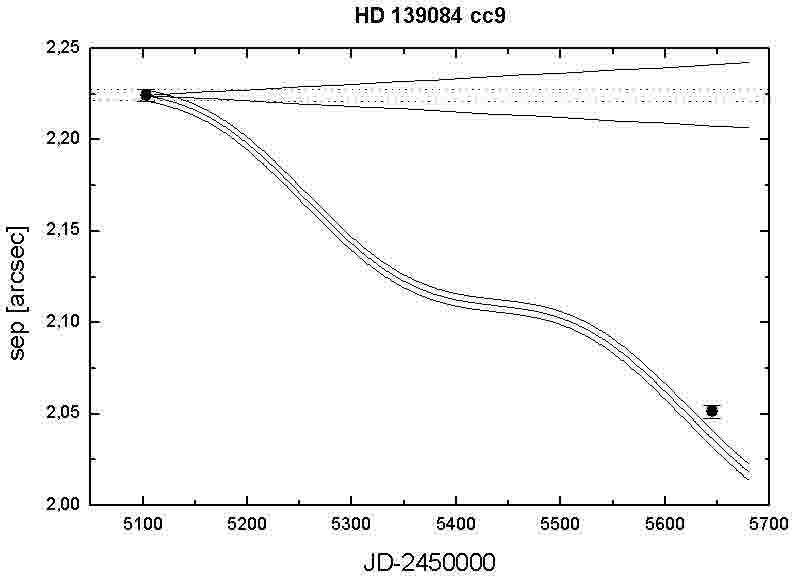}\includegraphics[angle=0]{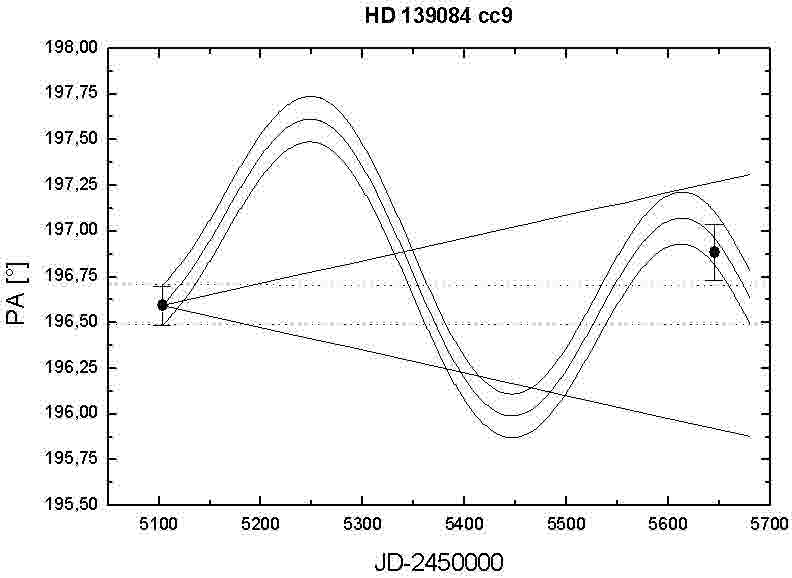}}
\end{figure*}
\clearpage

\begin{figure*}
\resizebox{\hsize}{!}{\includegraphics{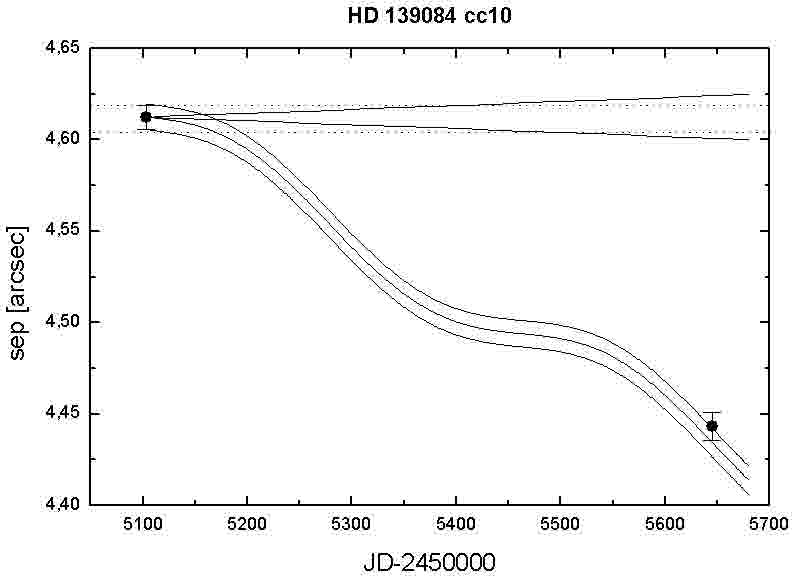}\includegraphics[angle=0]{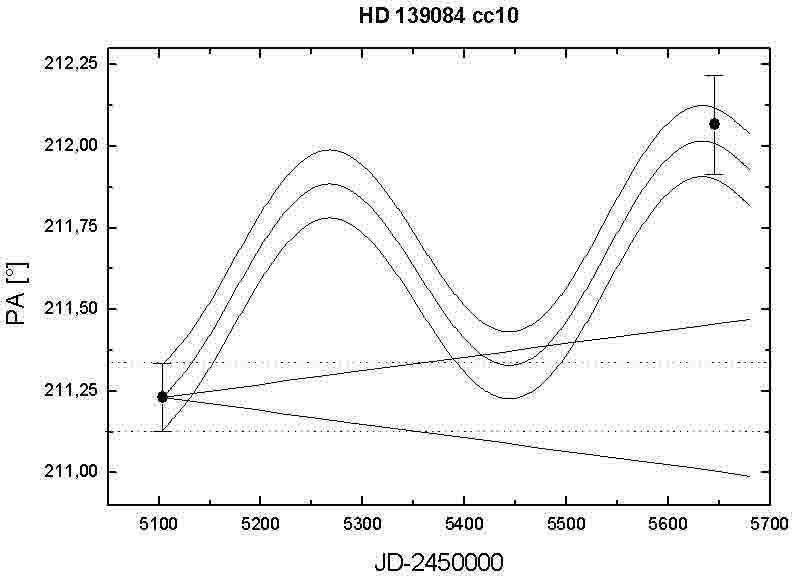}}
\resizebox{\hsize}{!}{\includegraphics{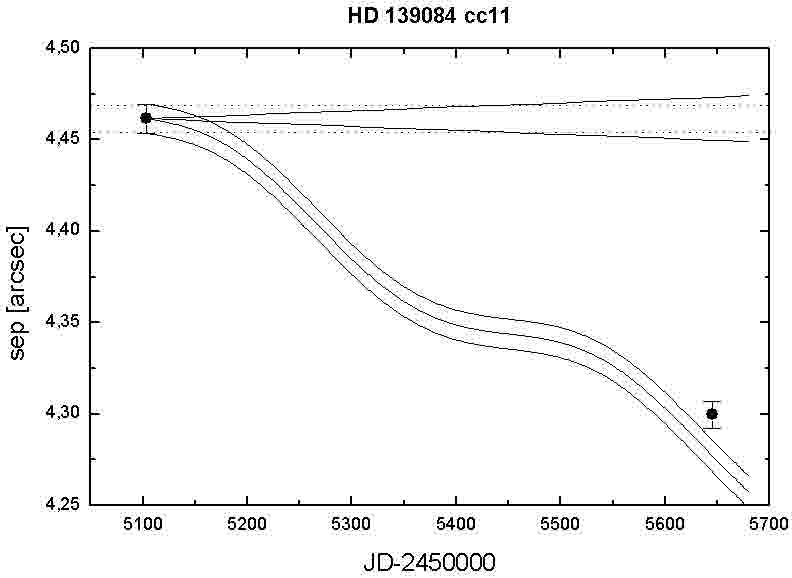}\includegraphics[angle=0]{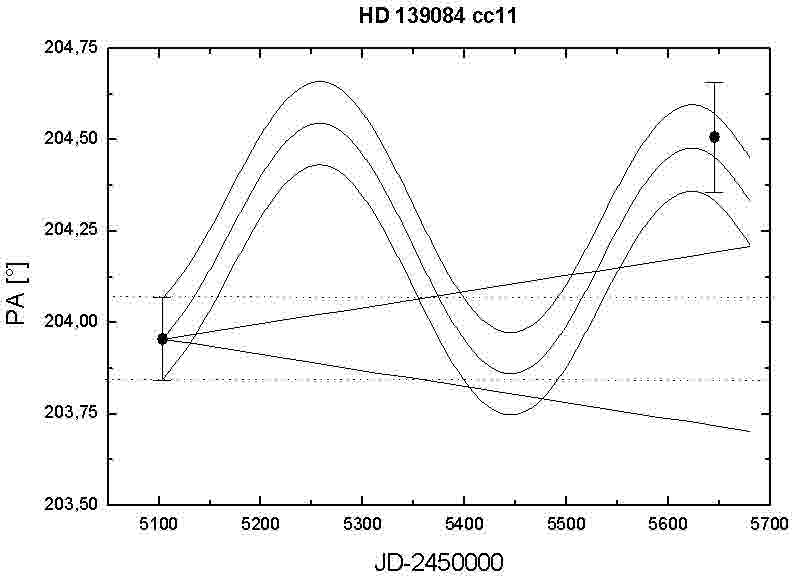}}
\resizebox{\hsize}{!}{\includegraphics{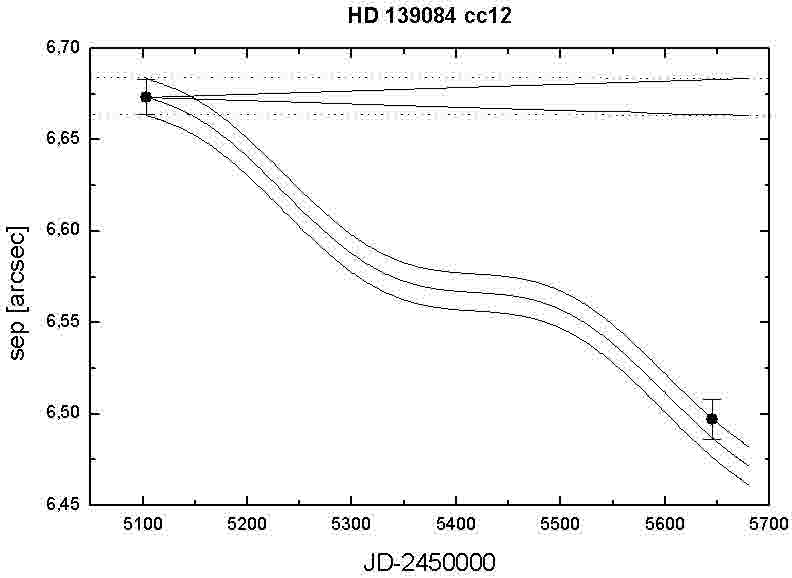}\includegraphics[angle=0]{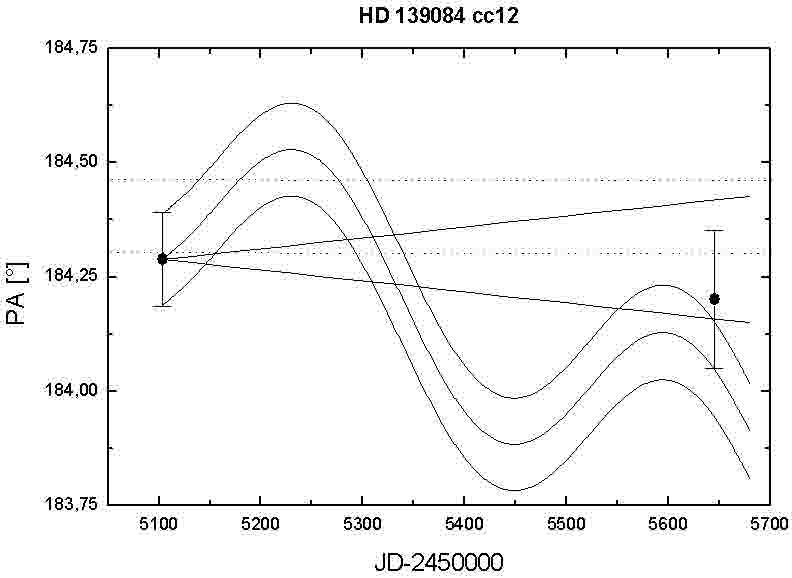}}
\end{figure*}
\clearpage

\end{document}